\documentclass[prb,aps,twocolumn,showpacs,amsmath,amssymb,floatfix,citeautoscript,10pt]{revtex4-1}
\usepackage{graphicx}
\usepackage{amsmath}
\usepackage{amssymb}
\usepackage{times}
\usepackage{epsfig,subfigure}
\usepackage{color}
\usepackage{float}
\usepackage{hyperref}

\newcommand {\vri} {{\bf R}} 
\newcommand {\vrri} {{\bf R}'} 
\newcommand {\sa} {\sigma} 
\newcommand {\ah} {\alpha} 
\newcommand {\vk} {{\bf k}}

\newcommand {\vq} {{\bf Q}}
\newcommand {\vqq} {{\bf q}}
\newcommand \vqaf {\vq_{\text{AF}}}
\newcommand \med[1] {\langle{#1}\rangle}

\hypersetup{
pdfauthor = {S\'eb,Chris and Cath},
pdftitle = {Spin liquid versus long range magnetic order in the frustrated body-centered tetragonal lattice. \today},
}

\begin{document}

\title{Spin liquid versus long range magnetic order in the frustrated body-centered tetragonal lattice}

\author{S\'{e}bastien Burdin}
\affiliation{Univ. Bordeaux, LOMA, UMR 5798, F-33400 Talence, France}
\affiliation{CNRS, LOMA, UMR 5798, F-33400 Talence, France}
\author{Christopher Thomas}
\affiliation{Institut de Physique Th\'eorique, CEA-Saclay, 91191 Gif-sur-Yvette, France}
\affiliation{International Institute of Physics, Universidade Federal do Rio Grande do Norte, 59078-400 Natal-RN, Brazil}
\affiliation{Instituto de F\'isica, UFRGS, 91501-970 Porto Alegre-RS, Brazil}
\author{Catherine P\'epin}
\affiliation{Institut de Physique Th\'eorique, CEA-Saclay, 91191 Gif-sur-Yvette, France}
\author{Alvaro Ferraz}
\affiliation{International Institute of Physics, Universidade Federal do Rio Grande do Norte, 59078-400 Natal-RN, Brazil}
\affiliation{Departamento de F\'isica Te\'orica e Experimental, Universidade Federal do Rio Grande do Norte,  59072-970 Natal-RN,Brazil}
\author{Claudine Lacroix}
\affiliation{Institut N\'eel, Universit\'e Grenoble-Alpes, F-38042 Grenoble, France}
\affiliation{Institut N\'eel, CNRS, F-38042 Grenoble, France}

\begin{abstract}
The quantum Heisenberg model is studied in the geometrically frustrated body-centered tetragonal lattice (BCT lattice) with antiferromagnetic interlayer coupling $J_1$ and intralayer first and second neighbor coupling $J_2$ and $J_3$. 
Using a fermionic representation of the spin $1/2$ operators, we introduce a variational method: each interaction term 
can be decoupled partially in the purely magnetic Weiss and in the spin-liquid (SL) mean-field channels. 
We find that the most stable variational solutions correspond to the three different possible long range magnetic orders that are respectively governed by $J_1$, $J_2$, and $J_3$. We show that magnetic and SL parameters do not coexist, and we characterize three different purely SL non-magnetic solutions that are variationally the second most stable states after the purely magnetic ones. 
The degeneracy lines separating the purely magnetic phases do not coincide with the ones separating the purely SL phases. 
This suggests that quantum fluctuations induced by the frustration between $J_1$-$J_2$-$J_3$ coupling 
should destroy magnetic orders and stabilize the formation of SL in large areas of parameters. The SL solution governed by $J_1$ breaks the lattice translation symmetry. This Modulated SL is associated to a commensurate ordering wave vector (1,1,1). 
Remarking that four different fits of experimental data on URu$_2$Si$_2$ locate this material with BCT lattice very close to  the degeneracy line between $J_1$ and $J_3$ but well inside the Modulated SL, we suggest that frustration might be a key ingredient for the formation of the Hidden order phase observed in this compound. Our results also underline possible  analogies between different families of correlated systems with BCT lattice, including unconventional superconductors. 
Also, the general variational method introduced here can be applied to any other system 
where interaction terms can be decoupled in two different mean-field channels. 
\end{abstract}

\date{\today}

\maketitle

\section{Introduction}

The body-centered tetragonal (BCT) lattice is one of the 14 three-dimensional lattice types~\cite{Kittelbook}. 
This standard crystalline structure is realized in several strongly correlated electron materials with unusual magnetic 
and transport properties. Among the heavy fermion systems~\cite{Stewart1984, Fulde2006}, 
different examples of materials with rare earth atoms on a BCT lattice have been intensively studied for the last decades: 
in URu$_2$Si$_2$, a still mysterious Hidden order (HO) phase was discovered in 1986, that appears below the critical temperature $T_{HO}\approx 17~{\rm K}$ close to a 
pressure-induced antiferromagnetic 
(AF) transition~\cite{Palstra1985, Mydosh2011}; in YbRh$_2$Si$_2$ and CeRu$_2$Si$_2$, non-Fermi liquid properties  are observed in the vicinity of AF quantum phase transitions, that are still poorly 
understood~\cite{Steglich2003,Steglich2009, Flouquet1988, Knafo2009}; 
CeCu$_2$Si$_2$ was the first (heavy fermion) material where unconventional superconductivity was discovered in 
1979 close to an AF transition~\cite{Steglich1979}; CePd$_2$Si$_2$ also exhibits unconventional superconductivity related to an AF transition~\cite{Lonzarich1998, Demuer2001}. Today, each one of these compounds can yet be considered as one entire field of research. 
It is noticeable that the link between AF ordering and unconventional superconductivity has also been suggested in other families of correlated materials with BCT symmetry: the cuprate superconductors, discovered in 1986 by Bednorz and 
Muller~\cite{Bednorz1986}, whose AF insulating parent compounds include La$_2$CuO$_4$ and 
Sr$_2$CuO$_2$Cl$_2$. In these cases, the AF order originates from the Cu atoms that form a BCT crystal. 
But the relevant physics there is mainly two-dimensional, the BCT structure being only involved in the formation of 
square-lattice layers of Cu atoms that order antiferromagnetically. 

The BCT lattice can also be considered as a prototype three-dimensional frustrated system. 
Important theoretical developments were made in the past years about the unconventional magnetic properties of the BCT lattice using a classical Heisenberg model. These works were motivated by the rich magnetic phase diagram of iron based materials like FePd possibly doped with Rh, with a main focus on the competition between ferromagnetic, AF,  and helical 
orders~\cite{Diep1989,Diep1989a,Rastelli1989,Quartu1998,Loison2000,Sorokin2011}. 
Tuning the interaction parameters made possible the description of different phases in the XY and Heisenberg models with thermal and quantum fluctuations. It was also shown that magnetic fluctuations as magnons excitations can help in the stability of long-range order. 

\begin{figure}[H]
\centering
\includegraphics[width=0.4\columnwidth]{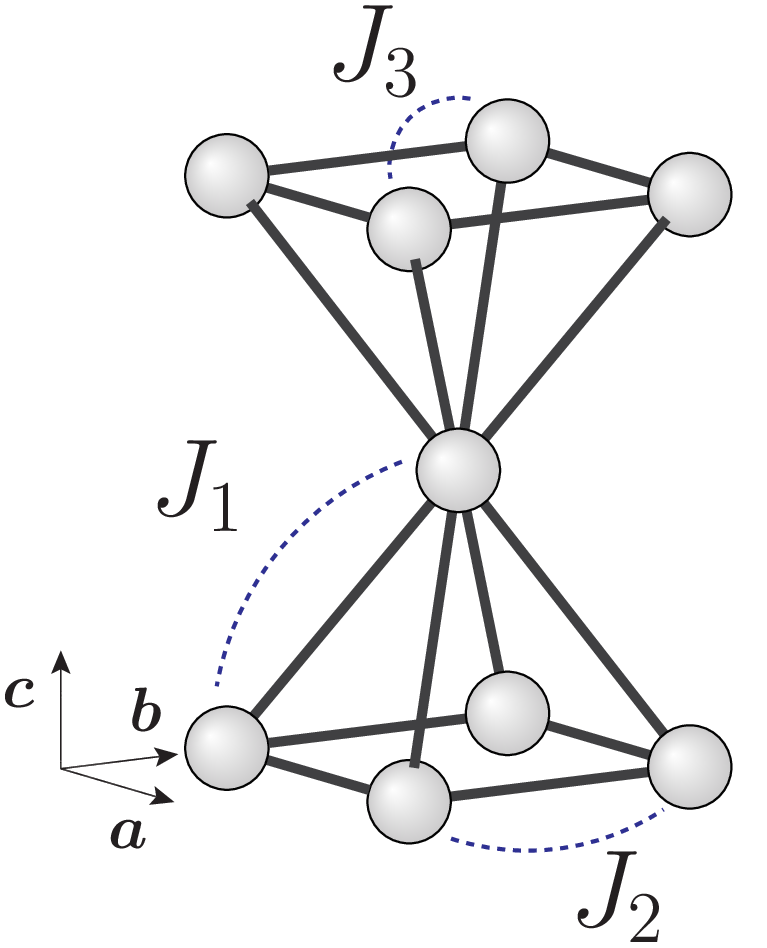}
\caption{\label{fig:bct} BCT lattice and the $J_1$, $J_2$, and $J_3$ interactions. 
Lattice constants are $a$ in the ${\bf a},{\bf b}$ directions and $c$ along ${\bf c}$.}
\end{figure}

In this paper, we analyze the ground states of a frustrated 
$J_1$-$J_2$-$J_3$ quantum Heisenberg model on a BCT lattice as 
illustrated by figure~\ref{fig:bct}. We are aware that a complete exact determination of its expected-to-be rich phase 
diagram would not be realistic and we thus need to do some approximations. 
Here, we introduce and use a variational mean-field method that allows to decouple the Heisenberg interaction terms partially 
in the standard Weiss and in the modulated spin liquid (MSL) channels. 
Since the MSL state has been initially introduced as a scenario for the HO state in 
URu$_2$Si$_2$~\cite{Pepin2011,Thomas2013, Montiel2013}, applications to this compound 
are considered as one motivation. However, the method which is developed here could be adapted to other correlated 
systems with BCT structure. 

The paper is organized as follows: section~\ref{section_Model} introduces the concept of MSL, the model, the mean-field decoupling, and the variational method. 
General results including phase diagrams are analyzed for $J_3=0$ and all $T$ in section~\ref{ResultsJ1J2}, and for 
$T=0$ and all $J_3$ in section~\ref{sec:j3mod}. 
We will see how geometric frustration that is intrinsic to the model may help stabilizing a MSL ordered ground state. 
Applications to real correlated materials with BCT lattice are discussed in section~\ref{Applications}.


\section{Model and method \label{section_Model}}

\subsection{The concept of modulated spin liquid (MSL)}
The expression spin liquid was originally introduced in 1976 by contrast with spin glasses, in order to describe the dynamical properties of a disordered spin system\cite{Aharony1976}. Nonetheless the concept of spin liquid within quantum Heisenberg models on frustrated geometries usually also refers to the Resonant Valence Bond (RVB) state proposed by Fazekas and Anderson in 1974 on the triangular lattice~\cite{Fazekas1974}. Later, 
Baskaran, Zou, and Anderson have proposed that RVB spin-liquid correlations could act like a magnetic glue for the 
Cooper pair formation in cuprate superconductors~\cite{Anderson1987, Baskaran1987, Rice1993, Wen1996}. 
Within this scenario, the AF N\'eel ordered state formed by the Cu square lattice layers in the insulating parent compounds is destabilized by charge fluctuations induced by doping on the O sites. The MSL scenario proposed for URu$_2$Si$_2$ was inspired by the spin-liquid scenario for cuprates. Even if the underlying BCT lattice is shared by these two families of systems, the microscopic physics in URu$_2$Si$_2$ is of course quite different and the long range orders invoke correlations in three-dimensions. 
Whether a system can have a true spin-liquid ground state or not has been a long standing issue, but some good evidences 
of possible spin-liquid ground states have been proposed for the Heisenberg model on frustrated lattices~\cite{Canals1998,Wen2002,Balents2010, Frustratedmagnetismbook}.  
It has also been observed from numerical calculations that spin-liquid disordered states can be very close in energy to 
dimer ordered states~\cite{Yan2011, Iqbal2013}. In general, spin dimer orders refer to bond orders that are characterized by  a given periodic pattern of disconnected dimers. 
The proposed MSL state can be thought of as a kind of spin dimer commensurate ordered state where two different dimers may be connected to a same site. Such a dimer ordered state may also be named valence bond 
crystal~\cite{Frustratedmagnetismbook} especially when it is characterized by bosonic triplet excitations. Here, we prefer use the name MSL because its magnetic excitations are deconfined Abrikosov fermions. 

In previous works, the competition between AF and MSL orders on a 
square lattice~\cite{Pepin2011} and on a BCT lattice~\cite{Thomas2013} was tuned phenomenologically by introducing 
two independent nearest neighbor coupling $J_{AF}$ and $J_{SL}$. 
Here, we study this competition as an intrinsic effect associated to the geometric frustration in the 
$J_1$-$J_2$-$J_3$ quantum Heisenberg model on the BCT lattice. We then introduce a variational method that allows 
to treat the system in a mean field approximation, where the interaction on each lattice bond can be decoupled in two channels, the magnetic and the spin liquid. The relative weight of each decoupling channel is determined by minimizing the free energy of the system.

\subsection{Model and method of calculation}
\subsubsection{The \texorpdfstring{$J_1$-$J_2$-$J_3$}{J1-J2-J3} model}
The $J_1$-$J_2$-$J_3$ model is defined by the following quantum Heisenberg Hamiltonian: 
\begin{align}
H&=\sum_{\langle {\bf R},{\bf R'}\rangle}\sum_{\sigma\sigma'}J_{{\bf RR}'}\chi_{{\bf R}\sigma}^{\dagger}\chi_{{\bf R}\sigma'}\chi_{{\bf R}'\sigma'}^{\dagger}\chi_{{\bf R}'\sigma}~, 
\label{eq:hheis}
\end{align}
where $\chi_{{\bf R}\sigma}^{\dagger}$ ($\chi_{{\bf R}\sigma}$) is the creation (annihilation) fermionic operator that represents quantum spins $1/2$, and satisfy the local constraints $\sum_{\sigma=\uparrow, \downarrow}\chi_{{\bf R}\sigma}^{\dagger}\chi_{{\bf R}\sigma}=1$.  The antiferromagnetic interactions $J_{{\bf RR}'}$ connects two sites 
${\bf R}$ and ${\bf R}'$ on a BCT lattice, and can take three possible values $J_1,~J_2,~J_3>0$, as indicated on 
figure~\ref{fig:bct}.

\subsubsection{Variational method}
In a very oversimplified classical mean field approach and considering the specific connectivity of the BCT lattice, 
we expect that competition between different Weiss mean fields may reveal degenerate frustrated ground states.
Hereafter, we go beyond this classical picture, and we introduce quantum correlation effects at a mean field level within a spin-liquid 
RVB-like decoupling on each bond. 
First we formally split the interaction term on each bond into two different contributions:   
\begin{align}
J_i&\equiv J_{i}^{\text{Weiss}}+J_{i}^{\text{SL}}\equiv J_i\cos^2(\alpha_i)+J_i\sin^2(\alpha_i)\,, 
\label{eq:DefJ1channels}
\end{align}
where $\alpha_1$, $\alpha_2$, and $\alpha_3$ are variational parameters. 
Hereafter, each interaction term will be treated within a mixed mean-field approximation on each bond: the mean-field 
decoupling will be made partially in the Weiss channel, and partially in SL channel. 
The extreme cases $\alpha_i=0$ and $\alpha_i=\pi/2$ correspond to a decoupling in the purely classical Weiss channel and in the purely SL channel, respectively. 
In the following, the three decoupling variational parameters $\alpha_{i}\in [0, \pi/2]$ will be determined self-consistently 
as functions of $J_1$, $J_2$ and $J_3$ in order to minimize the free energy of the system. 

\subsubsection{General mean-field decoupling}
Generalizing the procedure developed in Refs.~\onlinecite{Pepin2011,Thomas2013}, and invoking the variational 
splitting Eq.~\ref{eq:DefJ1channels}, the Heisenberg Hamiltonian (eq. \eqref{eq:hheis}) is decoupled for each bond 
${\bf RR}'$ using appropriated Hubbard-Stratonovich transformations as follows: 
\begin{eqnarray}
&J_{i}^{\text{Weiss}}&\sum_{\sigma\sigma'}\chi_{{\bf R}\sigma}^{\dagger}\chi_{{\bf R}\sigma'}\chi_{{\bf R}'\sigma'}^{\dagger}\chi_{{\bf R}'\sigma}\nonumber\\
&&\approx
J_{i}^{\text{Weiss}}\sum_{\sigma}
\left(\sigma m_{\vri}\chi_{\vrri\sa}^{\dag}\chi_{\vrri\sa}+\sigma m_{\vrri}\chi_{\vri\sa}^{\dag}\chi_{\vri\sa}\right)
\nonumber\\
&&~~-2J_{i}^{\text{Weiss}}m_{\bf R}m_{{\bf R}'}~, 
\label{ApproxmeanfieldWeiss}
\end{eqnarray}
where $m_{\bf R}$ is the local contribution from site ${\bf R}$ to the magnetic Weiss field, 
with $\sigma=\uparrow,\downarrow\equiv +,-$, and :
\begin{eqnarray}
&J_{i}^{\text{SL}}&\sum_{\sigma\sigma'}\chi_{{\bf R}\sigma}^{\dagger}\chi_{{\bf R}\sigma'}\chi_{{\bf R}'\sigma'}^{\dagger}\chi_{{\bf R}'\sigma}\nonumber\\
&&\approx
J_{i}^{\text{SL}}\sum_{\sigma}\left(\varphi_{{\bf RR}'}^\star\chi_{{\bf R}\sigma}^{\dagger}\chi_{{\bf R}'\sigma}+c.c.\right)
+
J_{i}^{\text{SL}}\vert \varphi_{{\bf RR}'}\vert^2~, \nonumber\\
&~&
\label{ApproxmeanfieldSL}
\end{eqnarray}
where $\varphi_{\vri\vrri}^\star\equiv\varphi_{\vrri\vri}$ denotes the spin-liquid field on the bond $\vri\vrri$. 
Hereafter, the Hubbard-Stratonovitch fields are replaced by their mean-field values, which are given by free energy 
saddle point conditions: 
\begin{eqnarray}
m_{\vri}&=&\frac{1}{2}\sum_{\sa}\sa\langle \chi_{\vri\sa}^{\dag}\chi_{\vri\sa}\rangle~, \\
\varphi_{\vri\vrri}&=&-\sum_{\sa}\med{\chi_{\vri\sa}^{\dag}\chi_{\vrri\sa}}~. 
\end{eqnarray}
The self-consistency of the mean-fields is established from the following mean-field Lagrangian: 
\begin{align}
{\cal L}&={\cal L}_1+{\cal L}_2+{\cal L}_3+\sum_{{\bf R}\sigma}\chi_{{\bf R}\sigma}^{\dagger}\left(\partial_\tau+\lambda_{\bf R}\right)\chi_{{\bf R}\sigma}-\sum_{\bf R}\lambda_{\bf R}~, \notag\\
&~ \label{Lagrangian}
\end{align}
with 
\begin{align}
{\cal L}_1&\equiv
\sum_{n}\sum_{\langle \vri\in P_n,\vrri\in P_{n+1}\rangle}\left[ 
J_1^{\text{SL}}\sum_{\sigma}\left(\varphi_{\vri\vrri}^\star\chi_{\vri\sa}^{\dag}\chi_{\vrri\sa}+c.c\right)
\right. \notag\\
&+
J_1^{\text{Weiss}}\sum_{\sigma}\left(\sigma 
m_{\vri}\chi_{\vrri\sa}^{\dag}\chi_{\vrri\sa}+\sigma m_{\vrri}\chi_{\vri\sa}^{\dag}\chi_{\vri\sa}\right)\notag\\
&+\left. J_1^{\text{SL}}\vert \varphi_{\vri\vrri}\vert^{2}-2J_{1}^{\text{Weiss}}m_{\bf R}m_{{\bf R}'}
\right]\,,  
\end{align}
where $P_n$ denotes sites of the planar layer $n$ oriented in the $a,b$ crystallographic directions indicated on figure~\ref{fig:bct}, 
and: 
\begin{align}
{\cal L}_2&\equiv
\sum_{n\sigma}\sum_{\langle \vri,\vrri\rangle\in P_n}\left[ 
J_2^{\text{SL}}\sum_{\sigma}\left(\varphi_{\vri\vrri}^\star\chi_{\vri\sa}^{\dag}\chi_{\vrri\sa}+c.c.\right)\right. \notag\\
&+
J_2^{\text{Weiss}}\sum_{\sigma}\left(\sigma m_{\vri}\chi_{\vrri\sa}^{\dag}\chi_{\vrri\sa}+\sigma m_{\vrri}\chi_{\vri\sa}^{\dag}\chi_{\vri\sa}\right)\notag\\
&+\left. J_2^{\text{SL}}\vert \varphi_{\vri\vrri}\vert^{2}-2J_2^{\text{Weiss}}m_{\bf R}m_{{\bf R}'}\right]\,,  
\notag\\
&~\notag\\
{\cal L}_3&\equiv
\sum_{n\sigma}\sum_{\langle \langle\vri,\vrri\rangle\rangle\in P_n}\left[ 
J_3^{\text{SL}}\sum_{\sigma}\left(\varphi_{\vri\vrri}^\star\chi_{\vri\sa}^{\dag}\chi_{\vrri\sa}+c.c.\right)\right. \notag\\
&+
J_3^{\text{Weiss}}\sum_{\sigma}\left(\sigma m_{\vri}\chi_{\vrri\sa}^{\dag}\chi_{\vrri\sa}+\sigma m_{\vrri}\chi_{\vri\sa}^{\dag}\chi_{\vri\sa}\right)\notag\\
&+\left. J_3^{\text{SL}}\vert \varphi_{\vri\vrri}\vert^{2}-2J_3^{\text{Weiss}}m_{\bf R}m_{{\bf R}'}\right]\,.  
\end{align}
In these expressions of ${\cal L}_1$, ${\cal L}_2$, and ${\cal L}_3$, the sums over bonds $\vri,\vrri$ are taken with the same 
connectivity as the couplings $J_1$, $J_2$, and $J_3$ respectively, which is indicated on figure~\ref{fig:bct}: in 
${\cal L}_1$ the bonds are nearest neighbors in two different planes $P_n$ and $P_{n+1}$, in ${\cal L}_2$ the bonds are nearest neighbors in the same plane $P_n$, and in ${\cal L}_3$ the bonds are second nearest neighbors in the same plane. 
The convention used in these notations is that each pair $\vri\vrri$ is summed only once. 

In the following, we will make some Ansatz for the mean-field parameters $m_{\vri}$ and $\varphi_{\vri\vrri}$, which will generalize the approach of Ref.~\onlinecite{Thomas2013}. This first requires to introduce space Fourier transforms and 
to use the momentum representation of the fermionic operators: 
\begin{eqnarray}
\chi_{{\bf k}\sigma}\equiv\frac{1}{\sqrt{N}}\sum_{\bf R}e^{-i{\bf k}\cdot{\bf R}}
\chi_{{\bf R}\sigma}\,, 
\end{eqnarray}
where $N$ is the number of lattice sites. The inverse relation is 
\begin{eqnarray}
\chi_{{\bf R}\sigma}\equiv\frac{1}{\sqrt{N}}\sum_{\vk\in\text{\bf BZ}_{\text{site}}^{\text{BCT}}}e^{i{\bf k}\cdot{\bf R}}\chi_{{\bf k}\sigma}\,. 
\end{eqnarray}
Here, $\text{\bf BZ}_{\text{site}}^{\text{BCT}}$ refers to the first Brillouin zone of the BCT lattice of sites. This precision will be useful later since other Brillouin zones will emerge from the dual lattices made of inplane and interplane bonds 
(see appendix~\ref{Apendixduallattice}). 
We define the mean-fields in reciprocal space as: 
\begin{eqnarray}
m_{\bf k}&\equiv&\frac{1}{\sqrt{N}}\sum_{\bf R}e^{-i{\bf k}\cdot{\bf R}}m_{\bf R}\,, 
\label{Defmk}\\
\varphi_{\bf q}^{1}&\equiv&\frac{e^{i\theta_{\bf q}}}{2\sqrt{N}}\sum_{n}
\sum_{\langle {\bf R}\in L_n, {\bf R}'\in L_{n+1}\rangle}e^{-i{\bf q}\cdot\left( \frac{{\bf R}+{\bf R}'}{2}\right)}\varphi_{\vri\vrri}^\star\,, \notag\\
&&~\label{Defvarphiq1}\\
\varphi_{\bf q}^{2}&\equiv&\frac{1}{\sqrt{2N}}\sum_{n}\sum_{\langle {\bf R}, {\bf R}'\rangle\in L_{n}}e^{-i{\bf q}\cdot\left( \frac{{\bf R}+{\bf R}'}{2}\right)}\varphi_{\vri\vrri}^\star\,,\label{Defvarphiq2}\\
\varphi_{\bf q}^{3}&\equiv&\frac{1}{\sqrt{2N}}\sum_{n}\sum_{\langle\langle {\bf R}, {\bf R}'\rangle\rangle\in L_{n}}e^{-i{\bf q}\cdot\left( \frac{{\bf R}+{\bf R}'}{2}\right)}\varphi_{\vri\vrri}^\star\,. \label{Defvarphiq3}
\end{eqnarray}
Here, a phase factor $\theta_{\bf q}\equiv {\bf q}\cdot{\bf R_0}$ is introduced for the interlayer spin-liquid field 
$\varphi_{\bf q}^{1}$ in order to fix the origin of the interplane bond lattice at real space position ${\bf R_0}\equiv ({\bf a}+{\bf b}+{\bf c})/4$. Such a global phase factors could be included arbitrarily for convenience to each mean-field. 
The site and bond dependence of the mean-fields can be recovered by the reciprocal Fourier relations: 
\begin{align}
m_{\bf R}\equiv\frac{1}{\sqrt{N}}\sum_{\vk\in\text{\bf BZ}_{\text{site}}^{\text{BCT}}}e^{i{\bf k}\cdot{\bf R}}m_{\bf k}\,,\label{TFmag}
\end{align}
and 
\begin{equation}
\varphi_{\vri\vrri}=
\begin{array}{|ll}
\varphi_{\vri\vrri}^i&\text{if }\vri\text{ and }\vrri\text{ are connected by }J_i\\
~~&\\
0&\text{else}~, 
\end{array}
\end{equation}
with
\begin{align}
\varphi_{\vrri\vri}^{1}&\equiv\frac{1}{2\sqrt{N}}\sum_{\vqq\in\text{\bf BZ}_{\text{bond}}^{1}}e^{i{\bf q}\cdot\left(\frac{{\bf R}+{\bf R}'}{2}\right){-i\theta_{\bf q}}}\varphi_{\bf q}^{1}\,, \label{TFphi1}\\ 
\varphi_{\vrri\vri}^{2}&\equiv\frac{1}{\sqrt{2N}}\sum_{\vqq\in\text{\bf BZ}_{\text{bond}}^{2}}e^{i{\bf q}\cdot\left( \frac{{\bf R}+{\bf R}'}{2}\right)}\varphi_{\bf q}^{2}\,,\label{TFphi2}\\
\varphi_{\vrri\vri}^{3}&\equiv\frac{1}{\sqrt{2N}}\sum_{\vqq\in\text{\bf BZ}_{\text{bond}}^{3}}e^{i{\bf q}\cdot\left( \frac{{\bf R}+{\bf R}'}{2}\right)}\varphi_{\bf q}^{3}\,, \label{TFphi3}
\end{align}
The different Brillouin zones emerging here from the dual lattices of bonds are defined and discussed in appendix~\ref{Apendixduallattice}. 
At this general stage, the number of mean-field variables that can be considered is still huge. Concerning the Weiss mean-fields $m_{{\bf k}}$, we consider here magnetic structures described by a single$-{\bf k}$
ordering wave-vector ${\bf Q}_{\text{AF}}$, excluding multi$-{\bf k}$ structures. 
Hereafter, we will generalize this classical mean-field approach by doing similar Ansatz for the bond spin liquid mean-fields. 

\subsubsection{Mean-field Ansatz}
Hereafter, the Weiss and spin liquid mean-fields are approximated using the following Ansatz: 
\begin{eqnarray}
m_{\bf R}&=&S_{{\bf Q}_{\text{AF}}}e^{i{\bf Q}_{\text{AF}}\cdot{\bf R}}~, \label{Ansatzm}\\
\varphi_{\vri\vrri}^{1}&=&\frac{1}{2}\Big[\Phi_1+ie^{i\vq\cdot\big(\frac{\vri+\vrri}{2}\big)}\Phi_{\vq}\Big]~, 
\label{Ansatzvarphi1}\\
\varphi_{\vri\vrri}^{2}&=&\Phi_2~, \label{Ansatzvarphi2}\\
\varphi_{\vri\vrri}^{3}&=&\Phi_3~. \label{Ansatzvarphi3}
\end{eqnarray}
Here, $S_{{\bf Q}_{\text{AF}}}$ is the staggered magnetization characterizing an AF order. The wave-vector ordering 
${\bf Q}_{\text{AF}}$ will be fixed by minimization of the spin-wave spectrum resulting from the Weiss field. 
The three fields $\Phi_1$, $\Phi_2$, and $\Phi_3$ correspond to the homogeneous parts of the spin liquid terms 
along the three kinds of bonds that are considered here. The emergence of three such homogeneous spin liquid fields is a natural BCT lattice generalization of the RVB decoupling introduced initially on triangular lattice~\cite{Fazekas1974}
 and later on a square lattice~\cite{Anderson1987, Baskaran1987}.  
The extra term $\Phi_{\bf Q}$ included in this Ansatz takes into account a possible spatial modulation of the spin liquid field. The specific choice of this spin-liquid modulation is motivated by previous work of Ref.~\onlinecite{Thomas2013}, where 
only the interplane spin-liquid term $\varphi_{\vri\vrri}^{1}$ was considered. This modulation is defined on the bond lattice 
by a wave-vector ${\bf Q}$, and it can lower the lattice translation symmetry. 
Invoking the momentum representation given by 
Eqs.~(\ref{Defmk}, \ref{Defvarphiq1}, \ref{Defvarphiq2}, \ref{Defvarphiq3}), the mean-field Ansatz 
Eqs.~(\ref{Ansatzm}, \ref{Ansatzvarphi1}, \ref{Ansatzvarphi2}, \ref{Ansatzvarphi3}) can be expressed as 
\begin{eqnarray}
m_\vk&=&S_{{\bf Q}_{\text{AF}}}\sqrt{N}\delta(\vk-{\bf Q}_{\text{AF}})~, \\
\varphi_{\bf q}^1&=&\Phi_1\sqrt{N}\delta({\bf q})+\Phi_{\bf Q}\sqrt{N}\delta({\bf q}-{\bf Q})~, \\
\varphi_{\bf q}^2&=&\Phi_2\sqrt{2N}\delta({\bf q})~,\\
\varphi_{\bf q}^3&=&\Phi_3\sqrt{2N}\delta({\bf q})~, 
\end{eqnarray}
where $\delta({\bf q})$ denotes the Dirac distribution. 
We also assume an homogeneous and constant Lagrange multiplier $\lambda_{\bf R}=\lambda_{0}$. 
Finally, within this mean-field Ansatz, the Lagrangian~(\ref{Lagrangian}) can be expressed explicitly in terms of the $\vk$-dependent fermions as 
\begin{align}
{\cal L}&=\sum_{\sa\vk}\chi_{\vk\sa}^{\dag}\Big(\partial_{\tau}+\lambda_{0}\Big)\chi_{\vk\sa}
+N\lambda_0
\notag \\ 
&\,+4J_{1}^{\text{SL}}\Phi_1\sum_{\sa\vk}\gamma_{1,\vk}\,\chi_{\vk\sa}^{\dag}\chi_{\vk\sa}
+NJ_1^{\text{SL}}\left(\vert\Phi_1\vert^2+\vert\Phi_{\vq}\vert^2\right)\notag\\
&\,+2J_{1}^{\text{SL}}e^{-i\theta_{\vq}}\Phi_{\vq}\sum_{\sa\vk}\gamma_{\vq,\vk}\left[\chi_{\vk\sa}^{\dag}\chi_{\vk+\vq,\sa}+c.c\right]\notag\\
&\,+2J_{2}^{\text{SL}}\Phi_2\sum_{\sa\vk}\gamma_{2,\vk}\,\chi_{\vk\sa}^{\dag}\chi_{\vk\sa}+2NJ_2^{\text{SL}}\vert \Phi_2\vert^2\notag \\ 
&\,+4J_{3}^{\text{SL}}\Phi_3\sum_{\sa\vk}\gamma_{3,\vk}\,\chi_{\vk\sa}^{\dag}\chi_{\vk\sa}+2NJ_3^{\text{SL}}\vert \Phi_3\vert^2\notag \\ 
&\,+\sum_{\sa\vk}\sa J_{\vq_{\text{AF}}}S_{\vq_{\text{AF}}}\chi_{\vk\sa}^{\dag}\chi_{\vk+\vq_{\text{AF}},\sa}-NJ_{\vq_{\text{AF}}}\left|S_{\vq_{\text{AF}}}\right|^2\,,
\label{MeanfieldLagrangian}
\end{align}
where the effective spin-wave dispersion is 
\begin{align}
J_{\vq_{\text{AF}}}&\equiv 8J_1^{\text{Weiss}}\gamma_{1,\vq_{\text{AF}}}+2J_2^{\text{Weiss}}\gamma_{2,\vq_{\text{AF}}}
+4J_3^{\text{Weiss}}\gamma_{3,\vq_{\text{AF}}}\,, \label{DefJQAF}
\end{align}
and the effective dispersions resulting from the spin-liquid decoupling are given by:
\begin{align}
\gamma_{1,\vk}&\equiv\cos{\left(\frac{k_xa}{2}\right)}\cos{\left(\frac{k_ya}{2}\right)}
\cos{\left(\frac{k_zc}{2}\right)}\,,\label{defgamma1k}\\
\gamma_{\vq,\vk}&\equiv\gamma_{1,\vk+\vq/2}\,,\label{defgammaQk}\\
\gamma_{2,\vk}&\equiv\cos{(k_xa)}+\cos{(k_ya)}\,,\label{defgamma2k}\\
\gamma_{3,\vk}&\equiv\cos{(k_xa)}\cos{(k_ya)}\,.\label{defgamma3k}
\end{align}
The values considered for $\vq_{\text{AF}}$ will be those that minimize the spin-wave dispersion 
$J_{\vq_{\text{AF}}}$. Hereafter, we will restrict the analysis to some specific 
modulating vectors ${\bf Q}$ in $\text{\bf BZ}_{\text{bond}}^{1}$ that are equivalent to $\vq_{\text{AF}}$ in 
$\text{\bf BZ}_{\text{site}}^{\text{BCT}}$ (definitions of the various Brillouin zones are discussed in  appendix~\ref{Apendixduallattice}). 
One key assumption that will be made in the following is that we will consider only breaking of symmetries that lead to 
commensurate order with doubling of the unit cell. This restrictive but realistic assumption has a crucial simplifying consequence: $2{\bf Q}$, ${\bf Q}+\vq_{\text{AF}}$, and $2\vq_{\text{AF}}$ are all equivalent to ${\bf 0}$. 
In the Lagrangian, 
the MSL and AF terms correlate fermions of momentum ${\bf k}$ with fermions of momenta ${\bf k}+{\bf Q}$ and 
${\bf k}+\vq_{\text{AF}}$. Therefore, there is no new harmonics generated by these interactions since the second 
harmonics would correlate momenta ${\bf k}+{\bf Q}$ and ${\bf k}+\vq_{\text{AF}}$ with ${\bf k}$. 
There could be more possible solutions obtained by considering non-equivalent ${\bf Q}$ and $\vq_{\text{AF}}$, but such solutions would correspond to a lowering of the lattice symmetry associated to a bigger unit cell made of more than two atoms. 

\subsubsection{Free energy functional}
Invoking the assumptions ${\bf Q}=\vq_{\text{AF}}$ and $2\vq_{\text{AF}}={\bf 0}$, the free energy can be expressed 
from the mean-field Lagrangian Eq.~(\ref{MeanfieldLagrangian}) as 
\begin{align}\label{eq:totalfj1j2j3}
F&(\alpha_1,\alpha_2,\alpha_3,\lambda_0,\Phi_{1},\Phi_{\vq},\Phi_2,\Phi_3,S_{\vq_{\text{AF}}})= \notag\\&
-\frac{k_B T}{2N}\sum_{\vk\in\text{\bf BZ}_{\text{site}}^{\text{BCT}}}\sum_{\sa, s=\pm}\ln{\left(1+e^{-\beta\Omega_{\vk}^{s}}\right)}-\lambda_{0}
-J_{\vqaf}\vert S_{\vq_{\text{AF}}}\vert^{2}\notag\\&
+J_{1}^{\text{SL}}\left(\vert\Phi_{1}\vert^{2}+\vert\Phi_{\vq}\vert^{2}\right)
+2J_{2}^{\text{SL}}\vert\Phi_{2}\vert^{2}+2J_{3}^{\text{SL}}\vert\Phi_{3}\vert^{2}\,. 
\end{align}
where the eigenenergies involved are given by
\begin{align}
&\Omega_{\vk}^{\pm}=\lambda_0+2J_2^{\text{SL}}\gamma_{2,\vk}\Phi_2+4J_3^{\text{SL}}\gamma_{3,\vk}\Phi_3\notag \\&
\pm 
\sqrt{
(J_{\vqaf})^2\vert S_{\vq_{\text{AF}}}\vert^2+16(J_1^{\text{SL}})^2\big[(\gamma_{1,\vk})^2\vert\Phi_{1}\vert^{2}+(\gamma_{\vq,\vk})^2\vert\Phi_{\vq}\vert^{2}\big]
}\,.
\end{align}
The explicit dependence of the free energy in terms of the variational decoupling fields $\alpha_1$, $\alpha_2$, and $\alpha_3$ is obtained 
from the definition Eq.~(\ref{eq:DefJ1channels}) by identifying $J_{i}^{\text{Weiss}}=J_i\cos^2{(\alpha_i)}$ and $J_{i}^{\text{SL}}=J_i\sin^2{(\alpha_i)}$. The Weiss field and spin liquid dispersion terms are given by 
Eqs.~(\ref{DefJQAF}, \ref{defgamma1k}, \ref{defgammaQk}, \ref{defgamma2k}, \ref{defgamma3k}). 
The mean-field and variational parameters correspond to the minima of the free energy. 

\section{Temperature phase diagram for \texorpdfstring{$J_3=0$}{J3=0} \label{ResultsJ1J2}}
Before analyzing the ground state of the $J_1$-$J_2$-$J_3$ model, we start with the 
simplified situation where $J_3=0$. In this section we are thus not concerned with the fields $\alpha_3$ and $\Phi_3$. 
Hereafter, we use the reduced notation ${\bf Q}\equiv (h,k,l)$ for the ordering wave-vectors 
${\bf Q}=2\pi(h/a,k/a,l/c)$. 
When stable, all magnetic phases are analyzed for the wave vectors $\vqaf^{\rm I}=(1,1,1)$ and $\vqaf^{\rm II}=(1/2,1/2,0)$, that correspond to 
the classical magnetic solution, i.e., with $\alpha_1=\alpha_2=0$. 
Experimental examples of these two kinds of classical N\'eel orders in BCT lattices are realized in the AF phases of URu$_2$Si$_2$ and cuprates for $\vqaf^{\rm I}$ and $\vqaf^{\rm II}$, respectively.

\subsection{Method of calculation for \texorpdfstring{$J_3=0$}{J3=0}}
In order to find the stable configuration for the $J_3=0$ case, we have to minimize the free energy 
functional Eq.~\eqref{eq:totalfj1j2j3}. It is first minimized as much as possible analytically as a function of the variational decoupling 
fields $\alpha_1$ and $\alpha_2$. To do this, we start by expressing the seven saddle point relations for 
$F(\alpha_1,\alpha_2,\lambda_0,\Phi_{1},\Phi_{\vq},\Phi_2,S_{\vq_{\text{AF}}})$. The resulting 
system of equations is detailed in appendix~\ref{sec:sadpoint}, and invokes several formal sums over momenta 
$\vk$. After non trivial but straightforward algebraic transformations this system can be rewritten as seven equations 
(\ref{eq:sa1}-\ref{eq:slam}) that involve five independents sums over $\vk$. Explicit expressions of these five sums are given in Eqs.~(\ref{DefAlambda0}-\ref{DefASQ}). 
The resolution of this system in general requires a numerical approach, but we also find some trivial solutions that may have a physical meaning. Hereafter we analyze more precisely the trivial solutions that are obtained when the variational decoupling parameters $\alpha_1$ and $\alpha_2$ take the extreme values $0$ or $\pi/2$. 
Physically, such trivial solutions correspond to decoupling the corresponding Heisenberg interaction term (with $J_1$ or with $J_2$) in a pure channel that is either Weiss or spin-liquid. 
\begin{table}
\begin{tabular}{|l|c|c|}
\hline
~~&$\alpha_1$ &$\alpha_2$\\
\hline
Case A&$0$ or $\pi/2$&$0$ or $\pi/2$\\
\hline
Case B&$0$ or $\pi/2$&free parameter\\
\hline
Case C&free parameter&$0$ or $\pi/2$\\
\hline
Case D&free parameter&free parameter\\
\hline
\end{tabular}
\caption{Characteristics of the four possible cases for the variational decoupling parameters $\alpha_1$ and $\alpha_2$. }
\label{Table1}
\end{table}
Hereafter, we analyze the possible solutions by considering the four different cases as defined in 
table~\ref{Table1}: 
\subsubsection{Case A} 
Here, we consider extreme values for $\alpha_1$ and $\alpha_2$ so that $\sin{(2\alpha_1)}=\sin{(2\alpha_2)}=0$. 
The saddle point equations~(\ref{eq:sa1}) and~(\ref{eq:sa2}) are thus trivially satisfied and we are left with 
Eqs.~(\ref{eq:sphi}-\ref{eq:slam}). Among these five remaining equations, some may also be satisfied trivially. 

The sub-case $(\alpha_1,\alpha_2)=(0,0)$ corresponds to the classical mean Weiss field approximation. The two possible antiferromagnetic ground states compete, characterized respectively by the ordering wave-vectors $\vqaf^{\rm I}$ and $\vqaf^{\rm II}$.  The corresponding temperature-coupling classical phase diagram is depicted in figure \ref{fig:j1j2magclas} as a function of the dimensionless parameters $T/J_1$ and $J_2/J_1$. The classical phase transition between these two kinds of AF orders is realized at finite temperature when $J_1=J_2$. 

The sub-case $(\alpha_1,\alpha_2)=(0,\pi/2)$ does not correspond to a physically realistic situation. Indeed, decoupling $J_1$ in the pure Weiss and $J_2$ in the pure spin liquid channels artificially bypasses the underlying frustration problem. 
Such a solution artificially induces ferromagnetic planes coupled antiferromagnetically among them: this is compatible with 
$J_1$ interaction. But the inplane spin liquid term $\Phi_2$ vanishes, leading to a $J_2-$independent unphysical 
solution. It will not be considered in the following. 

For $(\alpha_1,\alpha_2)=(\pi/2,0)$, the interplane SL field competes with the inplane magnetization Weiss field with $\vqaf^{\rm II}=(1/2,1/2,0)$. Here, since we restrict our analysis to commensurate orders with at most a doubling of the unit cell, we enforce $\Phi_{\bf Q}=0$. 
The phase diagram presents a pure homogeneous SL solution with only $\Phi_1$ non-zero 
for $J_2/J_1\lesssim0.3$, and a purely magnetic solution is recovered for $J_2/J_1>0.5$. But these two extreme situations 
are more appropriately described by taking $(\alpha_1,\alpha_2)$ equal to $(\pi/2,\pi/2)$ and $(0,0)$ respectively. 
A more interesting solution is found in the range $0.3\lesssim J_2/J_1\lesssim0.5$, where the homogeneous SL field $\Phi_1$ coexists with the inplane antiferromagnetic order. Nevertheless, in this regime of parameters, the 
magnetic order obtained with $(\alpha_1,\alpha_2)=(0,0)$ has a much lower energy. Therefore, in the following we will not consider the sub-case $(\alpha_1,\alpha_2)=(\pi/2,0)$.

The last trivial sub-case is $(\alpha_1,\alpha_2)=(\pi/2,\pi/2)$, corresponding to pure spin-liquid decoupling. Here, the interplane MSL phase competes with the intraplane SL phase. For $J_2<J_1$, the MSL is 
predominant. Comparing the values of the free energy obtained by considering three possible ordering 
wave vectors $(1,1,1)$, $(0,0,1)$, and $(1,0,0)$, we found that ${\bf Q}=(1,1,1)$ corresponds to the most stable 
MSL state. For $J_2\gtrsim J_1$ the intraplane SL takes place. The temperature-coupling phase diagram for this sub-case is depicted in figure~\ref{fig:j1j2sl}. Due to the lattice breaking of symmetry associated with the MSL field, the critical line 
$T_{\Phi_{\vq}}$ indicates a true phase transition that would survive beyond the mean-field. The other mean-field critical temperature $T_{\Phi_2}$ rather describes a crossover since the inplane spin-liquid field $\Phi_2$ here is homogeneous. 

\subsubsection{Case B}
In this case, the saddle point condition~(\ref{eq:sa2}) can be simplified as 
$\gamma_{2,\vqaf}|S_{\vqaf}|^2+|\Phi_2|^2=0$. Letting aside the trivial high temperature solution where both 
$S_{\vqaf}$ and $\Phi_2$ vanish, we consider here only the magnetic wave vector $\vqaf^{\rm II}$. 
Indeed Eq.~(\ref{defgamma2k}) gives 
$\gamma_{2,\vqaf^{\rm I}}>0$ but $\gamma_{2,\vqaf^{\rm II}}<0$. 
Here, as a consequence of relation~(\ref{eq:sa2}), the intraplane spin liquid field $\Phi_2$ is proportional to the local magnetization. 
Solving the remaining saddle point equations in the sub-case $\ah_1=0$, we find the numerical value 
$\sin^2(\ah_2)=0.675\pm 0.01$.
For the other sub-case, $\ah_1=\pi/2$, the pure MSL state has the most stable configuration until 
$J_2\lesssim 2J_1$, then the pure inplane solution with non zero $S_{\vqaf}$ and $\Phi_2$ is present for higher $J_2$.

\subsubsection{Case C} 
Here, excluding the extreme solutions for $\alpha_1$, the saddle point Eq.~(\ref{eq:sa1}) is simplified as 
$8\gamma_{1,\vqaf}\vert S_{{\bf Q}_{\rm AF}}\vert^2+\vert \Phi_1\vert^2+\vert \Phi_{\bf Q}\vert^2=0$. 
In this case, 
Eq.~(\ref{defgamma1k}) gives  
$\gamma_{1,\vqaf^{\rm I}}<0$ but $\gamma_{1,\vqaf^{\rm II}}>0$. Therefore, only the ordering wave-vector 
$\vqaf^{\rm I}$ is considered for the magnetic phase. 
The trivial solution with vanishing $S_{{\bf Q}_{\rm AF}}$, $\Phi_1$, and 
$\Phi_{\bf Q}$ is not considered here, and we thus focus on the phases where magnetic order coexist with interlayer spin-liquid fields. 
For the first sub-case $\alpha_2=0$ we naturally explore the situation with $\Phi_2=0$. 
But for $\alpha_2=\pi/2$ all the competing mean-fields may coexist. Typically, this sub-case has similarities with the pure 
spin-liquid one discussed above and illustrated by figure~\ref{fig:j1j2sl}: the parameter $J_1/J_2$ tunes the competition between the interlayer MSL order and the inplane SL. 
However, here, a non-zero MSL field must coexist with a non-zero local magnetization field. 

\subsubsection{Case D} 
This case is in principle the most general one since it corresponds to non extreme values of both $\alpha_1$ and 
$\alpha_2$. Nevertheless, this situation can not be realized and it would correspond to all mean-fields vanishing. 
Indeed, assuming that neither $\alpha_1$ nor $\alpha_2$ are extreme, the saddle point relations~(\ref{eq:sa1}) 
and~(\ref{eq:sa2}) give $8\gamma_{1,\vqaf}\vert S_{{\bf Q}_{\rm AF}}\vert^2+\vert \Phi_1\vert^2+\vert \Phi_{\bf Q}\vert^2=0$ and $\gamma_{2,\vqaf}|S_{\vqaf}|^2+|\Phi_2|^2=0$. Non zero solutions for the mean-field parameters would thus require an ordering wave-vector $\vqaf$ such that both $\gamma_{1,\vqaf}<0$ and 
$\gamma_{2,\vqaf}<0$. Since these two conditions cannot be realized simultaneously, neither by $\vqaf^{\rm I}$ nor 
by $\vqaf^{\rm II}$, we exclude case D from our study.

\subsection{Results for \texorpdfstring{$J_3=0$}{J3=0}}
\begin{figure}[H]
\centering
\includegraphics[width=0.65\columnwidth, angle=90]{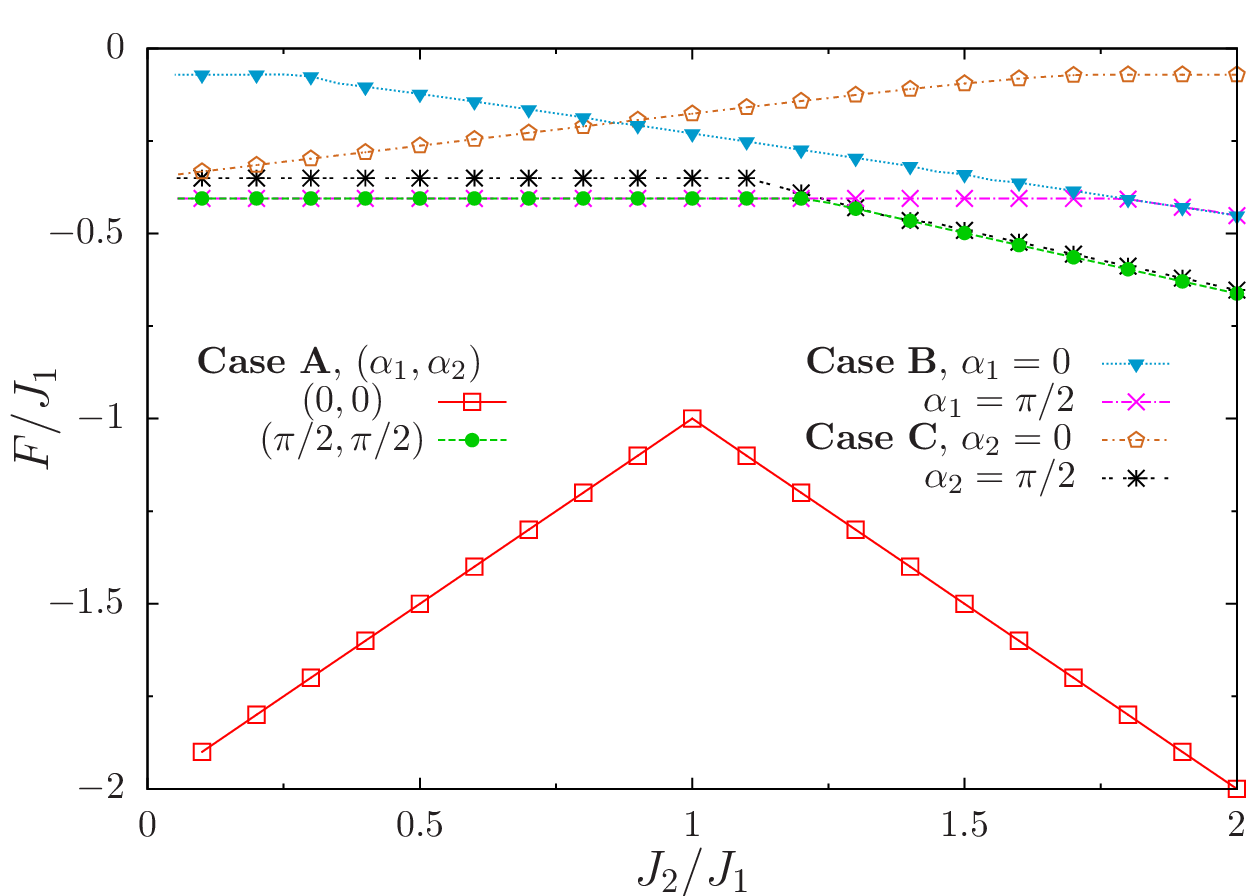}
\caption{\label{fig:fetot} Ground state energy of the model computed with $J_3=0$ as a function of $J_2/J_1$ for the various relevant cases discussed in this work and defined in table~\ref{Table1}. }
\end{figure}

All possible cases described above are studied by solving numerically the saddle point equations given in appendix \ref{sec:sadpoint}. 
We computed the free energy for each case, as functions of $J_2/J_1$ and $T/J_1$. For a sake of clarity, 
figure~\ref{fig:fetot} shows its evolution at $T=0$ only. The finite $T$ results are not presented here but they do not 
exhibit any extra free energy "crossing" between these cases. 

\begin{figure}[H]
\includegraphics[width=0.95\columnwidth]{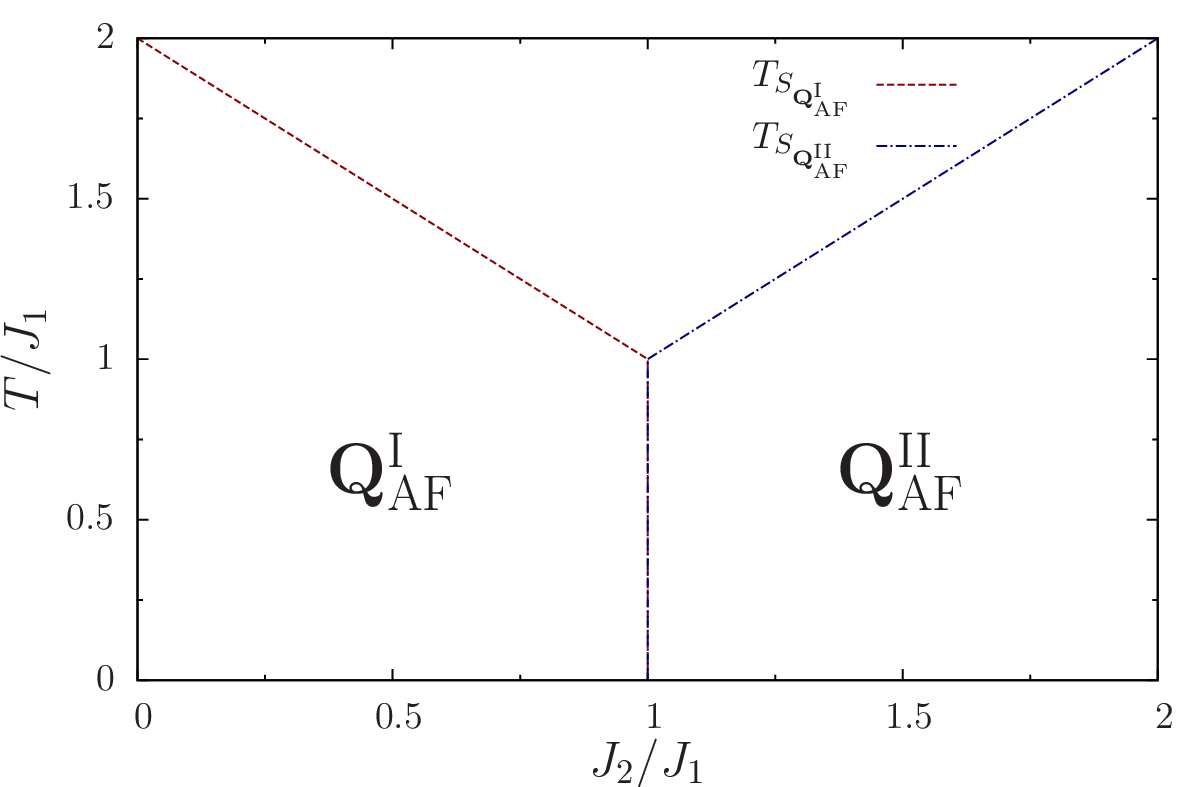}
\caption{Temperature-coupling phase diagram obtained with the purely magnetic configuration $\alpha_1=\alpha_2=0$ 
for $J_3=0$. The lines indicate the N\'eel ordering temperatures of the two magnetic orders corresponding to $\vqaf^{\rm I}=(1,1,1)$ and $\vqaf^{\rm II}=(1/2,1/2,0)$.
\label{fig:j1j2magclas}}
\end{figure}

\begin{figure}[H]
\includegraphics[width=0.95\columnwidth]{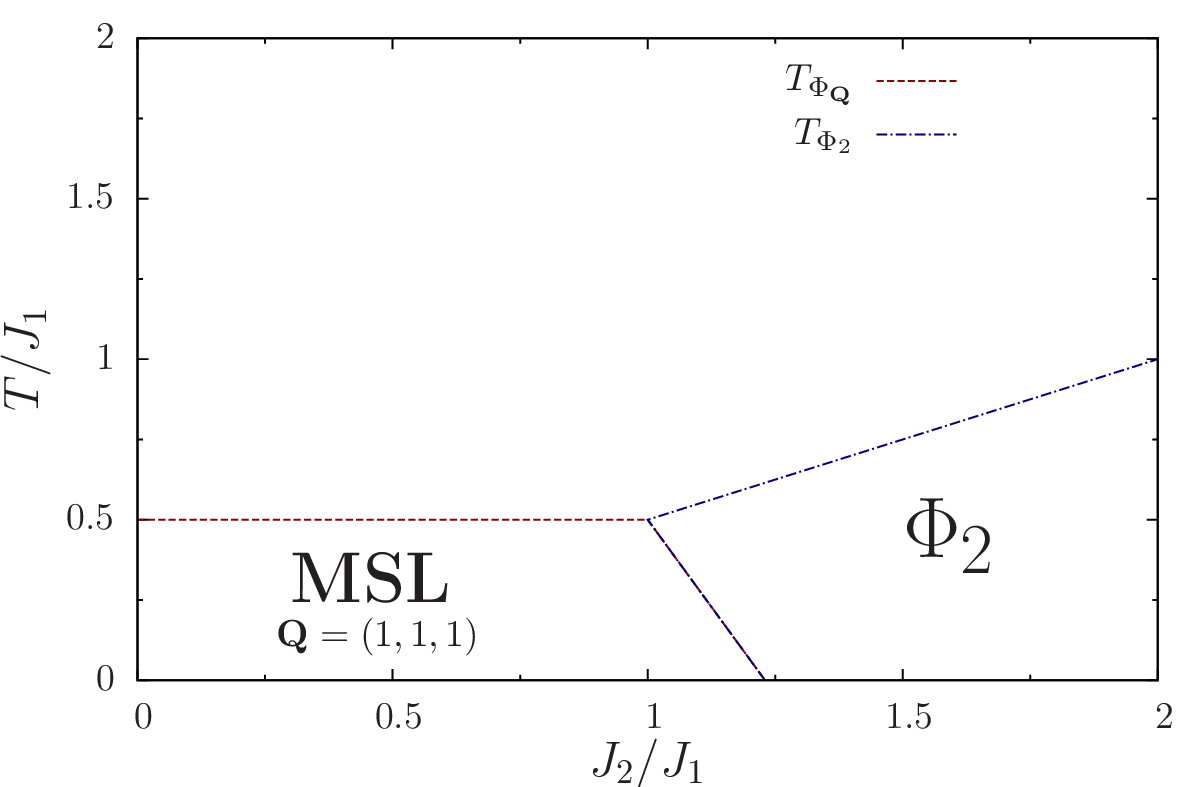}
\caption{Temperature-coupling phase diagram obtained with the purely spin-liquid decoupling channels  $\alpha_1=\alpha_2=\pi/2$ for $J_3=0$. The lines indicate the critical temperatures below which the corresponding 
mean-field $\Phi_1$, $\Phi_2$, and $\Phi_\vq$ are non-zero. Among these lines, 
$T_{\Phi_1}=T_{\Phi_\vq}$ is still expected to indicate a transition beyond the mean field because $\Phi_\vq$ is associated to a lattice symmetry breaking. $T_{\Phi_2}$ is expected to mark a crossover beyond the mean-field. 
\label{fig:j1j2sl}}
\end{figure}

The main result that emerges here from our variational approach for $J_3=0$ is the following: among all the considered cases, the classical purely AF mean-field solutions obtained with $\alpha_1=\alpha_2=0$ are always the most stable ones. 
The second most stable family of solutions are obtained with pure spin-liquid decoupling channels $\alpha_1=\alpha_2=\pi/2$. All the other combinations are found to be 
energetically less favorable. 
Here, we describe the two phase diagrams obtained for these two variational sub-cases. 
The temperature-coupling phase diagrams for both configurations $\alpha_1=\alpha_2=0$ and 
$\alpha_1=\alpha_2=\pi/2$ are shown in figures \ref{fig:j1j2magclas} and \ref{fig:j1j2sl} respectively.

\begin{figure}[H]
\includegraphics[width=0.95\columnwidth]{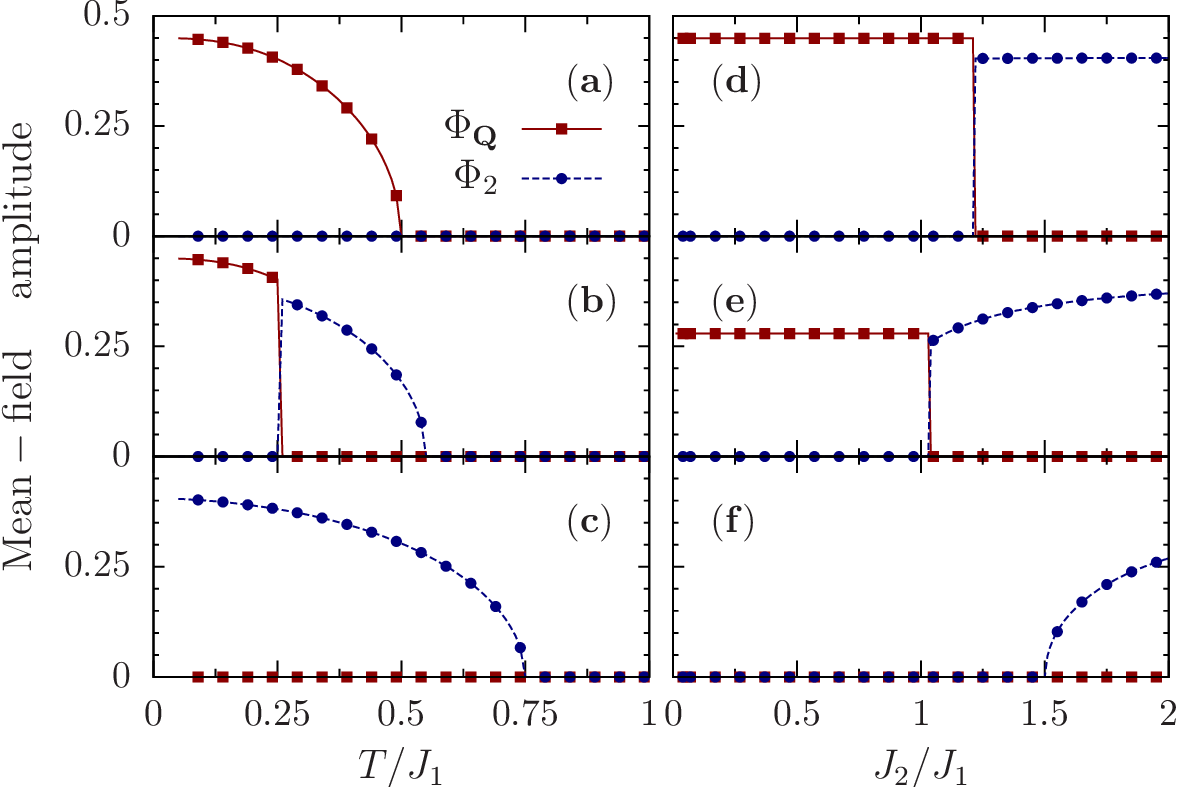}
\caption{Amplitude of the SL mean-field parameters $\Phi_{\bf Q}$ and $\Phi_1$ (red squares) and $\Phi_2$ (blue circles) 
computed for $J_3=0$ and $\alpha_1=\alpha_2=\pi/2$. Left: as a function of temperature for fixed 
$J_2/J_1=0.5$ (a), $1.1$ (b), $1.5$ (c). Right: as a function of $J_2/J_1$ for fixed temperature 
$T/J_1=0.05$ (d), $0.4$ (e), $0.75$ (f). 
With numerical accuracy we find $\Phi_1=\Phi_{\bf Q}$. 
\label{fig:j1j2meanfieldparameters}}
\end{figure}

While the purely AF solutions are the most stable the purely SL ones are energetically very close. 
Any mixed solution where both Weiss and SL mean-fields would coexist is found to be much less favorable 
and can also be excluded. 
Therefore, we can deduce that any fluctuation that would destabilize the AF order leave some room for stabilizing a 
pure SL phase. 
We also find that the SL parameters $\Phi_1=\Phi_{\bf Q}$ and $\Phi_2$ do not coexist, as illustrated by figure~\ref{fig:j1j2meanfieldparameters}. Depending on the value of $J_2/J_1$, there are three different kinds of temperature behaviors, 
corresponding to cases $a$, $b$, and $c$. 
Furthermore, we remark that the transition between the Modulated and the 
$\Phi_2-$dominated SL phases is characterized by a discontinuity of the corresponding mean-fields. This feature is in 
contrast with the continuous vanishing of these fields at the 
critical temperature separating the paramagnetic fully-decoupled phase from the SL ones. We thus conclude that the 
MSL transition is second order for $J_2<J_1$ and becomes first order for $J_2>J_1$. 
The transition temperature $T_{\Phi_{2}}$ is expected to indicate a crossover between the paramagnetic high $T$ and the SL low $T$ regimes when fluctuations beyond the mean-field approximation are included. Indeed, $\Phi_2$ is not associated to any breaking of symmetry. But we expect the transition at $T_{\Phi_{\bf Q}}$ to survive beyond the mean-field since the MSL phase is characterized by a breaking of lattice symmetry. 

An interesting feature also appears for the MSL solution: with a relatively high numerical accuracy 
the modulation field $\Phi_{\bf Q}$ is found to be always equal to the homogeneous field $\Phi_1$. 
Invoking the Ansatz Eq.~(\ref{Ansatzvarphi1}), this leads to a very extreme situation for the inter-layer field 
$\varphi_{\vri\vrri}^{1}=\frac{1}{2}[\Phi_1\pm\Phi_{\vq}]$ which vanishes on half of the bonds while it keeps 
the finite value $\Phi_1=\Phi_{\vq}$ on the other bonds. 
Introducing the probability $p_{\bf RR'}^{singlet}$ that a given bond ${\bf RR'}$ forms a singlet (see 
Appendix \ref{Apendixsingletprobability}), the formation of the MSL state can be interpreted here as follows: 
first, the interaction terms for all the inter-layer bonds such that $\vq\cdot ({\bf R}+{\bf R'})/2=\pi/2$ are effectively decoupled at the mean-field level, leading to a local probability $p_{\bf RR'}^{singlet}=1/4$ and a vanishing spin-spin correlations $\langle \vec{S}_{\bf R}\cdot\vec{S}_{\bf R'}\rangle=0$ . Then the spin-liquid with 
$\langle \vec{S}_{\bf R}\cdot\vec{S}_{\bf R'}\rangle\neq 0$ is formed on the other inter-layer bonds, with $\vq\cdot ({\bf R}+{\bf R'})/2=- \pi/2$, that remain effectively coupled. Using the numerical value 
$\Phi_1=\Phi_{\bf Q}\approx 0.45$ computed at $T=0$ in the MSL (see figure~\ref{fig:j1j2meanfieldparameters}), 
and invoking expression Eq.~(\ref{probasinglet}), we find that the singlet probability on these effectively 
coupled bonds is $p_{\bf RR'}^{singlet}\approx 0.60$. This value is, not surprisingly, higher than $1/4$, 
and it has to be compared with the value $ln(2)\approx 0.69$ that is predicted for a one-dimensional 
Heisenberg chain using exact methods like Bethe Ansatz~\cite{Bethe1931} or numerical renormalization 
technics~\cite{Schollwock2005}. We may thus interprete the MSL as a crystal of interacting 
filaments formed by the connected effectively coupled bonds. In this picture, spin excitations are deconfined fermions moving along the filaments. This may generalize the usual concept of valence bond crystal where localized spin $1$ excitations 
correspond to confined fermions.

\section{Mean-field ground state of the \texorpdfstring{$J_1$-$J_2$-$J_3$ model}{J1-J2-J3 model}}\label{sec:j3mod}
Here we analyze the ground state of the $J_1$-$J_2$-$J_3$ model within the mean-field Ansatz described above. 
In the previous section it was shown that for $J_3=0$ the low temperature most stable configuration is  
obtained by choosing purely magnetic Weiss mean-field decoupling channels. The second most stable solution corresponds to 
the purely spin liquid decoupling channels. Here, we assume that this result can be extended to the decoupling of the 
intraplane next nearest neighbor interaction $J_3$. We therefore assume that $\alpha_3$ can take only the extreme values 
$0$ or $\pi/2$.

\begin{figure}[H]
\centering
\includegraphics[width=0.95\columnwidth]{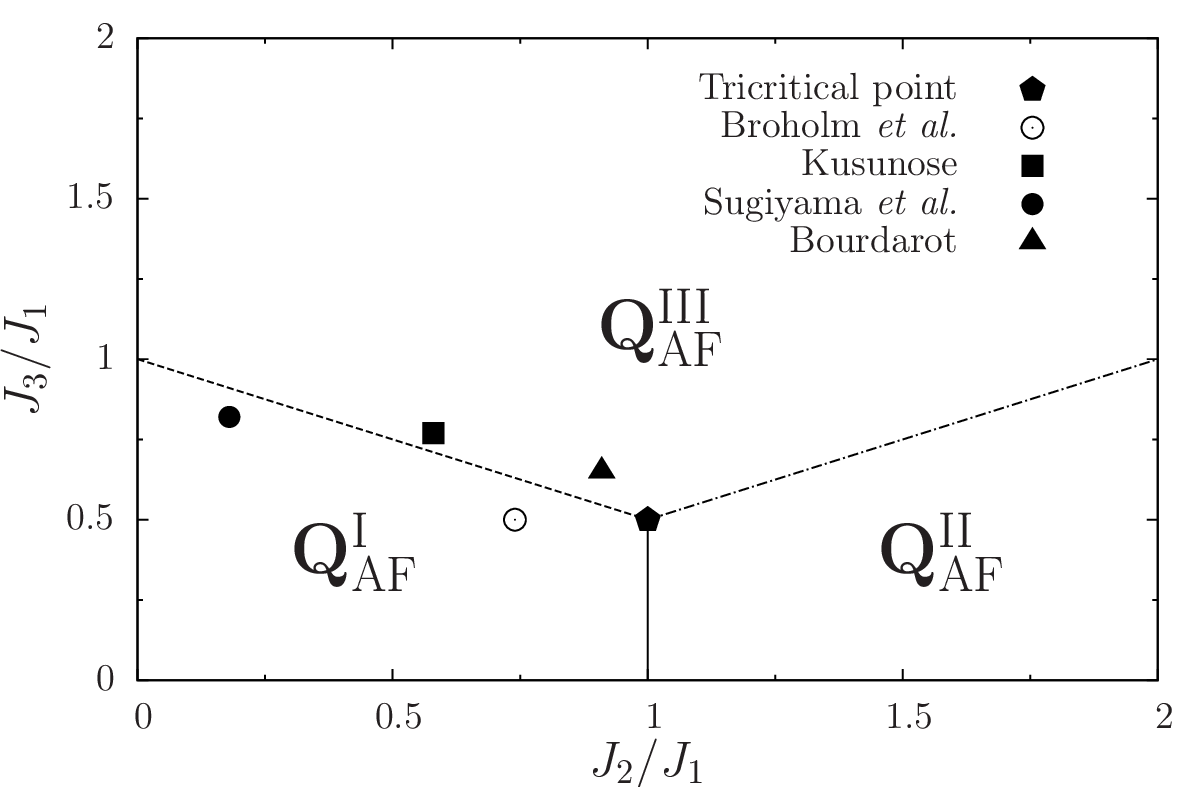}
\caption{\label{fig:magj1j2j3} 
Phase diagram characterizing the ground state of the $J_1$-$J_2$-$J_3$ model obtained within 
the pure Weiss mean-field decoupling channels $\alpha_1=\alpha_2=\alpha_3=0$. 
Three different magnetic orders are found, characterized by the wave-vectors $\vqaf^{\rm I}$, $\vqaf^{\rm II}$ and $\vqaf^{\rm III}$. 
We name {\it magnetic tricritical point} the highly degenerate point corresponding to the crossing of the three critical lines. Additionally, we include four points obtained from various fits of inelastic neutron scattering (INS) data on URu$_2$Si$2$: 
from Broholm {\it et al.}~\cite{Broholm1991}, Kusunose {\it et al.}~\cite{Kusunose2012}, Sugiyama {\it et al.}~\cite{Sugiyama1990} and Bourdarot~\cite{BourdarotHDR}. }
\end{figure}

\begin{figure}[H]
\centering
\includegraphics[width=0.95\columnwidth]{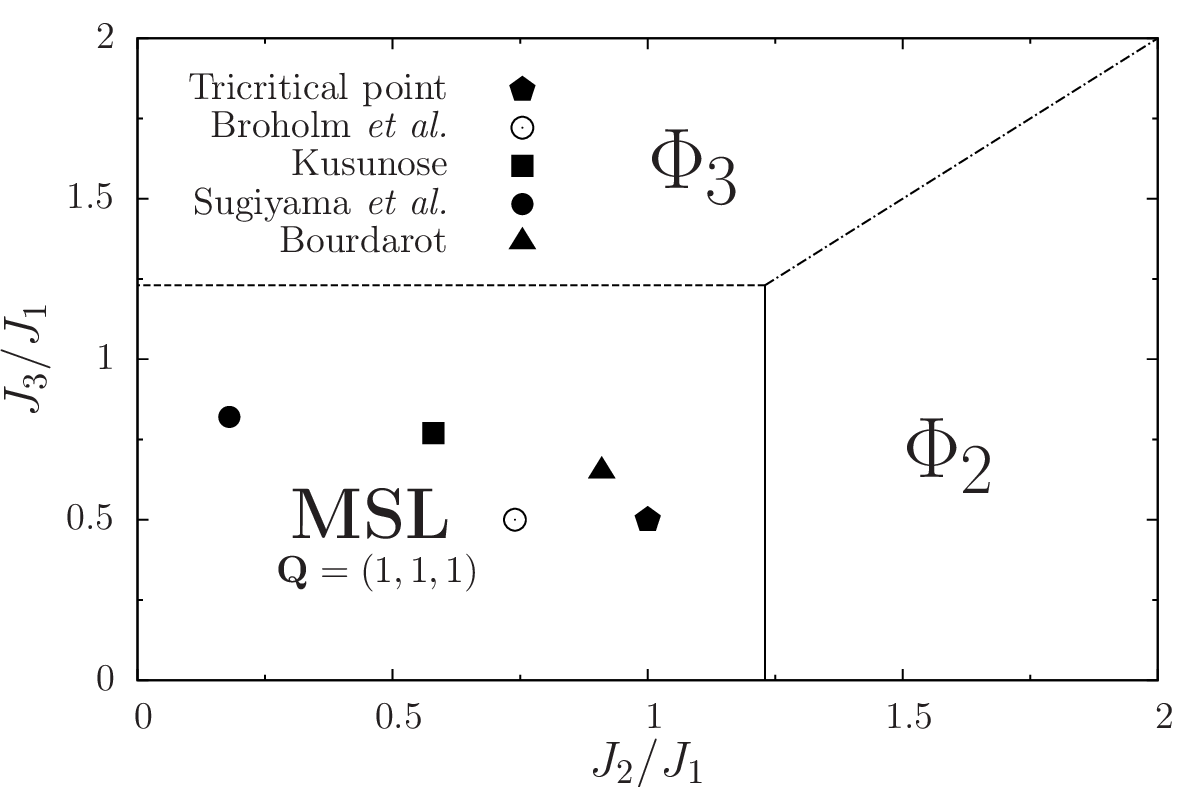}
\caption{\label{fig:slj1j2j3} Phase diagram characterizing the ground state of the $J_1$-$J_2$-$J_3$ model obtained within 
the pure spin-liquid mean-field decoupling channels $\alpha_1=\alpha_2=\alpha_3=\pi/2$. 
The MSL phase corresponds to finite $\Phi_1$ and $\Phi_\vq$. The two other spin-liquid phases correspond to a vanishing $\Phi_\vq$ and finite values of the nearest and next nearest neighbor inplane spin liquid fields $\Phi_2$ and $\Phi_3$ respectively. Among the three critical lines depicted here, only the ones indicating the MSL phase would still correspond to a 
transition when considering fluctuations beyond the mean-field approximation.  
The magnetic tricritical point is defined as the highly degenerate point in the purely magnetic phase 
diagram. The additional points obtained from INS data are included here with the same notations as 
in figure~\ref{fig:magj1j2j3}. }
\end{figure}

Solving numerically the two extreme cases, we find that, at the mean-field level, the classical magnetic solution with $\alpha_1=\alpha_2=\alpha_3=0$ is the most stable variational configuration. 
The resulting ground state phase diagram is presented in figure~\ref{fig:magj1j2j3} as a function of the dimensionless parameters $J_2/J_1$ and $J_3/J_1$. Three possible ordering wave-vectors are obtained, 
$\vq_{\text{AF}}^{\rm I}=(1,1,1)$, $\vq_{\text{AF}}^{\rm II}=(1/2,1/2,0)$, or $\vq_{\text{AF}}^{\rm III}=(1/2,0,0)$, that correspond to the three different regimes where the Weiss field can be dominated by $J_1$, $J_2$, or $J_3$ respectively. 
A highly degenerate point is found for $J_1=J_3=2J_2$, that we name {\it magnetic tricritical point}. 

Figure \ref{fig:slj1j2j3} depicts the phase diagram obtained within a purely spin-liquid mean-field decoupling $\alpha_1=\alpha_2=\alpha_3=\pi/2$. 
At the mean-field level we find three different phases, that are characterized by finite values of 
$\Phi_\vq$, $\Phi_2$, or $\Phi_3$. Beyond the mean-field, we expect that only the critical line defining finite $\Phi_\vq$ would still correspond to a phase transition, associated with a translation symmetry breaking.  
We remark that the MSL solution that we obtain corresponds to $\Phi_1=\Phi_\vq$, and it corresponds to the 
formation of a crystal of connected filaments as described above.

The position of the magnetic tricritical  point is also indicated in the pure spin liquid phase diagram, figure \ref{fig:slj1j2j3}. 
It is very surprising to see that this point which is highly degenerate from a Weiss mean-field perspective turns to be located well inside the MSL phase. 
Several earlier works have been dedicated to the characterization of the magnetic ground state of a frustrated Heisenberg 
model on a square lattice~\cite{Oitmaa1996,Sushkov2001,Kalz2011}, that can be realized here for $J_1=0$. It was shown that quantum fluctuations can stabilize a non magnetic spin-liquid phase between the antiferromagnetic phases $\vq_{\text{AF}}^{\rm II}$ and $\vq_{\text{AF}}^{\rm III}$. 
For this reason, we expect that huge quantum fluctuations of the Weiss mean-field should occur around all the critical 
lines separating the three possible phases $\vq_{\text{AF}}^{\rm I}$, $\vq_{\text{AF}}^{\rm II}$, and $\vq_{\text{AF}}^{\rm III}$. 
The position of the magnetic tricritical point inside the MSL phase suggests that the fluctuations of the MSL mean-field should be much less critical. Therefore, we expect that fluctuations beyond the mean-field will destabilize the magnetic solutions around all their degeneracy lines. We believe that the MSL mean-field solution should be more robust for all the regions that are sufficiently far from the MSL critical line. This is the case, for example, of the area around the magnetic tricritical point. 

\section{Discussion and applications to materials with BCT structure \label{Applications}}
\subsection{Relevance for Hidden order in \texorpdfstring{URu$_2$Si$_2$}{URu2Si2}}
The HO phase in URu$_2$Si$_2$ cannot be explained by the formation of too tiny local magnetic moments. Nevertheless, 
there are strong experimental evidences that the thermodynamic anomaly measured at the 
transition~\cite{Maple1986} has a magnetic origin. 
For example, the HO phase is characterized by a peak revealed by 
Inelastic Neutron Scattering (INS) at the commensurate wave-vector $\vq_{\text{AF}}=(1,0,0)$ in reduced 
notation~\cite{Broholm1987, Villaume2008, Bourdarot2010}. 
This wave-vector is surprisingly identical to the one that describes the pressure-induced AF phase of this compound. In the BCT structure, this AF order represents a ferromagnetic correlation in the ${ \bf a},{ \bf  b}$ directions (see Fig.~\ref{fig:bct}), with  antiferromagnetic correlations between nearest 
$({\bf  a},{\bf  b})$ planes. Recently, it was proposed that a quantum modulated spin liquid (MSL) phase could be stabilized by frustration 
and explain the origin of the hidden order phase in URu$_2$Si$_2$~\cite{Pepin2011,Thomas2013, Montiel2013}. 
A phase with a similar order as the MSL has also been proposed in terms of unconventional spin-orbital density 
wave~\cite{Riseborough2012, Das2014, Oppeneer2011}, where the order parameter characterizes a spatial commensurate 
modulation of the intersite hybridization between $5f$ states.
 
The first Heisenberg model on a BCT lattice that was proposed for URu$_2$Si$_2$ was introduced by Broholm {\it et al.}~\cite{Broholm1991}, trying to fit INS data in terms of spin density wave (SDW) excitations from an AF ground state. 
As we will see further, the resulting SDW model obtained by Broholm corresponds to a highly frustrated situation. 
The SDW scenario has later been contradicted by several other experiments. Nonetheless, the classical version of a $J_1$-$J_2$-$J_3$ Heisenberg model has been proposed by 
Sugiyama {\it et al.}~\cite{Sugiyama1990,Kim2003}  as a frustration scenario to explain the cascade of metamagnetic-like transitions and magnetization plateaux that are observed in URu$_2$Si$_2$. More recently, INS data analysis was invoked by Kusunose who proposed a competition between multipolar and AF Ising-like orders as a scenario for the HO-AF pressure-induced transition~\cite{Kusunose2012}. 
Bourdarot also recently proposed numerical values for $J_1$, $J_2$ and $J_3$ in order to fit his INS data~\cite{BourdarotHDR}. 

We are aware that modeling URu$_2$Si$_2$ with the present $J_1$-$J_2$-$J_3$ quantum Heisenberg model may 
constitute a very crude approximation with respect to several aspects: for example, the real system is metallic, and also, 
local 5f electronic states require an Ising-like  highly anisotropic multiplet description. Nevertheless, the numerous previous attempts 
to fit 
INS data using effective SDW dispersions make it worth checking where the fitted parameter would locate URu$_2$Si$_2$ on 
the mean-field phase diagrams we analyzed here. 

Hereafter, we use four different fits of various INS datas: the original fit introduced by Broholm {\it et al.} in Ref.~\cite{Broholm1991}, the fit introduced more recently by Kusunose~\cite{Kusunose2012} from Broholm's datas, 
the fit of INS datas from Sugiyama {\it et al.}~\cite{Sugiyama1990,Kim2003}, and the one from Bourdarot's data~\cite{BourdarotHDR}. 
These fits invoke not only $J_1$-$J_2$-$J_3$ terms but also up to seven Heisenberg-like interaction parameters 
in the BCT structure. Neglecting these extra parameters, we extracted the numerical values of $J_1$, $J_2$ and $J_3$ 
provided by each fit. The corresponding dimensionless pairs of ratios $J_2/J_1$ and $J_3/J_1$ thus provide 
specific points in the phase diagrams as indicated on figures~\ref{fig:magj1j2j3} and~\ref{fig:slj1j2j3}. 
The absolute numerical values of $J_1$, $J_2$ and $J_3$ that were provided by these four different fits do not coincide. This quantitative difference between fits is easily understandable: different experimental INS data were involved, and different extra fitting parameters were also involved, that we have not considered here. 
Nevertheless, it is remarkable that the four different fits all provide antiferromagnetic values for $J_1$, $J_2$, and $J_3$. 
Furthermore, the most interesting observation is the following: all of these different fits locate URu$_2$Si$_2$ 
in the very close vicinity of the transition line separating the two ordered states $\vqaf^{\rm I}$ and $\vqaf^{\rm III}$, 
as indicated on figure~\ref{fig:magj1j2j3}. 
We thus expect frustration to be very important as also noticed by Sugiyama {\it et al.}~\cite{Sugiyama1990,Kim2003}, and 
spin fluctuations may destabilize the magnetically ordered phase. Considering now the spin-liquid phase diagram on 
figure~\ref{fig:slj1j2j3}, we find that the four points that correspond to the different fits of INS data are all 
located well inside the MSL phase. 

This observation together with the analysis presented here suggest the MSL scenario as an alternative to the geometrical frustration problem that seems to prevent URu$_2$Si$_2$ from forming an AF order: the pressure induced HO-AF transition which is observed in this compound at low temperature could be mostly controlled by the tuning of $J_3/J_1$. 
At ambient pressure, quantum fluctuations are too strong and only the MSL state is realized. Applying pressure pushes the system away from the critical line, reducing the fluctuations and thus stabilizing the AF state with wave-vector ordering $\vqaf^{\rm I}$. 

We are aware that this scenario should be completed by including charge fluctuation effects and by taking into account the precise $5f$ local multiplet structure at the origin of the magnetic ordering. 
We believe that the concept of spatially modulated highly entangled state which emerges here from frustration would survive when adding such sophistications to the $J_1$-$J_2$-$J_3$ model. 

\subsection{Relevance for other systems}
Here we considered a model with only localized spins. But we know from previous works on cuprates and heavy-fermions 
that charge fluctuations play a crucial role in destabilizing antiferromagnetic states. 

In the context of cuprates, the AF N\'eel ordered phase of the insulating parent compounds corresponds to $\vqaf^{\rm II}$. The spin liquid phase introduced by Anderson 
{\it et al.}~\cite{Fazekas1974,Anderson1987, Baskaran1987, Rice1993, Wen1996} corresponds to the homogeneous spin liquid phase with $\Phi_2$ non zero. 
The relation between the 
spin-liquid field and the superconducting order parameter has been discussed by Wen, Lee 
{\it et al.} in terms of gauge transformations~\cite{Lee2006,Wenbook}. 
These gauge transformations are based on particle-hole transformations on the 
fermionic operators $\chi_{{\bf R}\sigma}$ that preserve the physical starting Heisenberg spin Hamiltonian but transform the spin-liquid fields into superconducting pairing terms. 

Doping may be introduced more generally on the full $J_1$-$J_2$-$J_3$ model. 
In heavy fermions, we know that the localized quasiparticle states are associated with the $f$-electrons. These localized degrees of freedom directly related to magnetism are usually distinguishable from the itinerant charge degrees of freedom. Indeed, in Ce and Yb compounds, delocalized modes emerge from light conduction electrons; in actinides they emerge from the duality of the $5f$ orbitals that have a partialy Mott-delocalized sector. In cuprates, such a localized spin - delocalized charge scenario cannot be clearly done. Especialy at low doping, the adaption of the present spin-fermion model for cuprates should include the physics of the Mott transition. Therefore, doping the $J_1$-$J_2$-$J_3$ model should be realized appropriately in various maners adapted to each experimental motivation: typically, within Kondo+Heisenberg, $t$-$J$, or multi-orbital Hubbard models. 

Inspired by the previous works of Wen, Lee {\it et al.}, we expect that the resulting charge fluctuations would strengthen the spin fluctuations and weaken the magnetically ordered phases that are predicted from a classical Heisenberg 
$J_1$-$J_2$-$J_3$ model. 
In turn, the spin-liquid phases are expected to remain stable, associated to superconducting instabilities. 
Invoking this general scenario, we predict that the symmetries of the resulting superconducting order parameters will 
result from the point group symmetries of the spin-liquids. 
This scenario may be tested first with the superconducting instability observed in URu$_2$Si$_2$ inside the HO phase. 
More generally, this scenario also generalizes to 3D systems the spin-fluctuation pairing mechanism that was proposed for cuprates. Here, the link between the BCT lattice structure and the superconducting order parameter is natural. 
This spin-liquid mechanism driven by frustration on the BCT lattice may also be tested for the heavy-fermion 
superconductors CeRu$_2$Si$_2$ and CePd$_2$Si$_2$, but in these systems valence fluctuation effects need to be 
carefully included. 

Appart from superconductivity, we may also question whereas there is a connexion between HO in URu$_2$Si$_2$ 
and the magnetic-field induced non-fermi liquid properties observed in YbRh$_2$Si$_2$. Indeed, this very 
unconventional heavy-fermion compound has a magnetically ordered ground state at ambiant pressure but the 
associated local moment is relatively small. This suggests that frustration on the BCT lattice may be analyzed together 
with Kondo screening in this system.

\section{Conclusion}\label{sec:con}
To summarize, we studied the frustrated $J_1$-$J_2$-$J_3$ quantum Heisenberg Hamiltonian in the BCT lattice using 
mean field approximations. Introducing variational parameters $\alpha_i$, each intersite interaction is decoupled 
in the Weiss and the spin liquid channels. 
Our first observation corresponds to the fact that variationally the 
interactions always prefer a pure channel. Indeed, any intermediate value of $\alpha_i$ corresponds to a higher free energy than the one obtained with decoupling parameters $\alpha_i=0$ (pure Weiss) or $\pi/2$ 
(pure spin-liquid). 

Studying the model at $J_3=0$ for all temperatures $T$ and at $T=0$ for all values of coupling $J_i$, we find that the most stable variational solution corresponds to the purely magnetically ordered ones. Nevertheless, we also analyze and characterize the purely SL solutions that are the second most stable ones. Three possible different magnetically ordered phases emerge at low $T$, characterized by the ordering wave-vectors $(1,1,1)$, $(1/2,1/2,0)$, and $(1/2,0,0)$ 
that respectively correspond to the three different regimes dominated by $J_1$, $J_2$, or $J_3$. 
Similarly, three different SL phases are also identified, the one dominated by $J_1$ corresponding to a non-homogeneous MSL state with commensurate ordering wave-vectors $(1,1,1)$, that is expected to survive beyond the mean-field. 
We also remarked that other variational solutions, including MSL states with a different wavevector $(0,0,1)$ or $(1,0,0)$ 
and mixed states with $\alpha_i$ non extreme, are energetically above but not so far from the three pure SL ones that are analyzed here. Fluctuations might stabilize some of these solutions as well. 

Whilst the purely magnetically ordered phases are the most stable at the mean-field level, we expect fluctuations to be strong in the vicinity of the degeneracy lines separating the different ordering wave-vectors. 
It is very interesting to notice that the analogous degeneracy lines obtained for the three different SL solutions do not coincide with the ones obtained for the magnetically ordered solutions. We thus conclude that fluctuations should open 
a large area of parameters where magnetic orders are destroyed, favoring the stabilization of SL phases. 

Surprisingly, when considering four different fits of experimental INS datas on URu$_2$Si$_2$, we find in each case 
that this compound is close to the degeneracy line separating the $(1,1,1)$ and $(1/2,0,0)$ antiferromagnetic orders. 
We also find that, when considering the SL solutions, each of these four fits locates URu$_2$Si$_2$ well inside the 
MSL phase. This result suggests that fluctuations and frustration between $J_1$ and $J_3$ coupling should play a crucial role in the HO-AF transition that is induced by pressure at low $T$ in this compound. The possible formation of a spatially modulated highly entangled state analogous to the MSL, emerging from frustration and fluctuations, could provide a key ingredient in the realization of the Hidden order phase. 

The scenario presented here is very general and could be adapted and applied to study doped correlated systems with BCT structure, including possibly unconventionnal superconductors. In these cases, the inclusion of charge fluctuations in the model are necessary and have to be done carefully since they might also play a crucial direct role for the superconducting instabilities.  
Finally, the variational method that we introduced here could also be used for other models where a two-body interaction term can be decoupled in two different mean-field channels.

\begin{acknowledgments}
We acknowledge the financial support of Capes-Cofecub Ph 743-12. CT is {\it bolsista Capes}.
This research was also supported in part by the Brazilian Ministry of Science, Technology and Innovation (MCTI) and the Conselho Nacional de Desenvolvimento Cient\'ifico e Tecnol\'ogico (CNPq). Research carried out with the aid of the Computer System of High Performance of the International Institute of Physics-UFRN, Natal, Brazil. The authors are gratefull to Fr\'ed\'eric Bourdarot for usefull discussions. 
\end{acknowledgments}

\appendix
\section{Brillouin zones for the dual (bond) lattice \label{Apendixduallattice}}
The choice of the phase of the modulation $+$ or $-$ on a given bond $\vri\vrri$ in Eq.~(\ref{Ansatzvarphi1}) is of course not unique. At this stage we could not go further by considering the system in its whole generality. Motivated by experimental applications to URu$_2$Si$_2$, we may thus assume that the order parameter $\Phi_{\bf Q}$ lowers the lattice translation symmetry from BCT to tetragonal. 
This translation symmetry breaking corresponds to a doubling of the lattice unit cell, and may as well be characterized by various point group symmetry breaking. Indeed, the spin-liquid field $\varphi_{\vri\vrri}$ is defined on the dual (bond) lattice. Each of these possible point group symmetry breaking results from a non isotropic distribution of the 
phase modulation $+$ or $-$ on the bonds neighboring a given lattice site. Different possible orders 
belong to the same tetragonal lattice group but break different point group symmetries. 
It is remarkable that a MSL order can equivalently be characterized by a point group symmetry or by 
an ordering wave-vector ${\bf Q}$ belonging to the reciprocal space of the dual lattice. On the other side, the AF 
order is characterized by a wave-vector ${\bf Q}_{\text{AF}}$ that belongs to the first Brillouin zone of the BCT lattice of sites, $\text{\bf BZ}_{\text{site}}^{\text{BCT}}$. 
We will thus later consider three other Brillouin zones, denoted 
$\text{\bf BZ}_{\text{bond}}^{1}$, $\text{\bf BZ}_{\text{bond}}^{2}$, and $\text{\bf BZ}_{\text{bond}}^{3}$, that 
correspond to the first Brillouin zones of the bonds connected with the couplings $J_1$, $J_2$, and $J_3$ respectively (see figure~\ref{fig:bct}). Note that $\text{\bf BZ}_{\text{bond}}^{2}$ and $\text{\bf BZ}_{\text{bond}}^{3}$ look like two-dimensional Brillouin zones since the couplings $J_2$ and $J_3$ are inplane. 
We remark here that the present formalism at this stage can be applied to study both two-dimensional magnetism in compounds like cuprates where $J_1\approx 0$ and three-dimensional magnetism in compounds like URu$_2$Si$_2$ for which $J_1$ drives the AF order. 
Since the BCT lattice has four times more bonds of kind $1$ than sites, it appears that 
$\text{\bf BZ}_{\text{site}}^{\text{BCT}}$ is four times smaller than $\text{\bf BZ}_{\text{bond}}^{1}$. As a result, 
different wave-vectors ${\bf Q}$ in $\text{\bf BZ}_{\text{bond}}^{1}$ characterizing different MSL bond orders, can be equivalent with each other from the AF point of view. For example, the ordering wave-vector 
${\bf Q}_{\text{AF}}^{\rm I}$ can be equivalently chosen to be $(1,1,1)$, $(1,0,0)$ or $(0,0,1)$ when 
characterizing AF ordered phase on the BCT lattice. 
But these three vectors characterize three different MSL ordered states. 
A detailed analysis is given in Ref.~\cite{Thomas2013}, comparing the free energy of these three possible MSL ordered states. 
It was found that, when the degeneracy was left, $(1,1,1)$ characterized the MSL state with the lowest free energy. 
Therefore, we choose to consider in this article only the results obtained with the modulation wave-vector 
${\bf Q}=(1,1,1)$. 
Finally, note that the prefactors $1/2\sqrt{N}$ and $1/\sqrt{2N}$ in the Fourier transform 
relations~(\ref{TFphi1}, \ref{TFphi2},\ref{TFphi3}) are related to the number of sites or bonds which are relevant for each field: the BCT lattice considered here has $N$ sites, $4N$ bonds connected by $J_1$, and $2N+2N$ bonds connected by $J_2$ and $J_3$.

\section{Bond singlet probabilities\label{Apendixsingletprobability}}
Keeping in mind that the fermionic operators $\chi_{\bf R\sigma}$ represent quantum spin $1/2$, each interaction term on 
a bond ${\bf RR'}$ in the $J_1$-$J_2$-$J_3$ Hamiltonian~(\ref{eq:hheis}) can be identified to an antiferromagnetic 
Heisenberg interaction: 
\begin{eqnarray}
\sum_{\sigma\sigma'}\chi_{{\bf R}\sigma}^{\dagger}\chi_{{\bf R}\sigma'}\chi_{{\bf R}'\sigma'}^{\dagger}\chi_{{\bf R}'\sigma}
=\frac{1}{2}+2\vec{S}_{\bf R}\cdot\vec{S}_{\bf R'}~, 
\end{eqnarray}
where $\vec{S}_{\bf R}$ and $\vec{S}_{\bf R'}$ are quantum spin $1/2$ on sites ${\bf R}$ and ${\bf R'}$. 
Integrating formally the other sites degrees of freedom of the many-body state characterizing the lattice, 
each local bond ${\bf RR'}$ can be characterized by a probability $p_{\bf RR'}^{singlet}$ to be in a singlet state. 
Invoking standard quantum spin algebra, we find the very general identity: 
\begin{eqnarray}
p_{\bf RR'}^{singlet}=\frac{1}{4}-\langle \vec{S}_{\bf R}\cdot\vec{S}_{\bf R'}\rangle~. 
\end{eqnarray}
Introducing the variational parameter $\alpha_i$ that is appropriate to the bond ${\bf RR'}$ as defined by 
Eq.~(\ref{eq:DefJ1channels}), and invoking the mean-field approximation decoupling in Weiss and spin-liquid channels 
as defined by Eqs.~(\ref{ApproxmeanfieldWeiss}) and~(\ref{ApproxmeanfieldSL}), we find the average 
\begin{eqnarray}
\sum_{\sigma\sigma'}\langle\chi_{{\bf R}\sigma}^{\dagger}\chi_{{\bf R}\sigma'}\chi_{{\bf R}'\sigma'}^{\dagger}\chi_{{\bf R}'\sigma}\rangle
&=&
2m_{\bf R}m_{\bf R'}\cos^2(\alpha_i)\nonumber\\
&&-\vert\varphi_{\bf RR'}\vert^2\sin^2(\alpha_i)~. 
\end{eqnarray}
Finally, within the variational mean-field approximation, the probability that a bond 
${\bf RR'}$ forms a singlet state is given by: 
\begin{eqnarray}
p_{\bf RR'}^{singlet}=\frac{1}{2}-m_{\bf R}m_{\bf R'}\cos^2(\alpha_i)
+\frac{\vert\varphi_{\bf RR'}\vert^2}{2}\sin^2(\alpha_i)~,  \nonumber\\
~~
\label{probasinglet}
\end{eqnarray}
where the kind of bond $i=1,~2$ or $3$ is defined on figure~\ref{fig:bct}.

\section{Saddle point equations for \texorpdfstring{$J_3=0$}{J3=0} \label{sec:sadpoint}}
Using expression~(\ref{eq:totalfj1j2j3}) with $\alpha_3=\Phi_3=J_3=0$, the seven saddle point equations for the 
free energy functional $F(\alpha_1,\alpha_2,\lambda_0,\Phi_{1},\Phi_{\vq},\Phi_2,S_{\vq_{\text{AF}}})$ are obtained 
from the following partial derivative expressions: 
\begin{align}
\frac{\partial F}{\partial \ah_1}&=2J_1\cos{\ah_1}\sin{\ah_1}\Bigg\{16\sum_{\vk}\frac{f(\Omega_{\vk}^+)-f(\Omega_{\vk}^-)}{\Delta \Omega_{\vk}}\notag\\
&\times\Big[2J_1\sin^2{\ah_1}(
\vert\Phi_{1}\gamma_{1,\vk}\vert^2+\vert\Phi_{\vq}\gamma_{1,\vk+\vq/2}\vert^2)-J_{\vqaf}\gamma_{1,\vqaf}|S_{\vqaf}|^2\Big]\notag\\
&+|\Phi_1|^2+|\Phi_{\vq}|^2+8\gamma_{1,\vqaf}|S_{\vqaf}|^2\Bigg\}\,,
\end{align}
\begin{align}
\frac{\partial F}{\partial \ah_2}&=4J_2\sin{\ah_2}\cos{\ah_2}\Bigg\{\sum_{\vk}\Big\{\big[f(\Omega_{\vk}^+)+f(\Omega_{\vk}^-)\big]|\Phi_2|\gamma_{2,\vk}\notag \\
&-2\Big[\frac{f(\Omega_{\vk}^+)-f(\Omega_{\vk}^-)}{\Delta \Omega_{\vk}}\Big]J_{\vqaf}\gamma_{2,\vqaf}|S_{\vqaf}|^2\Big\}\notag\\
&+|\Phi_2|^2+\gamma_{2,\vqaf}|S_{\vqaf}|^2\Bigg\}\,,
\end{align}
\begin{align}
\frac{\partial F}{\partial \Phi_2}&=2J_2\sin^2{\ah_2}\notag\\
&\times\Bigg\{\sum_{\vk}\Big[f(\Omega_{\vk}^+)+f(\Omega_{\vk}^-)\Big]\gamma_{2,\vk}+2|\Phi_2|\Bigg\}\,,
\end{align}
\begin{align}
\frac{\partial F}{\partial \Phi_1}&=2J_1\sin^2{\ah_1}|\Phi_1|\notag\\
&\times\Bigg\{16\sum_{\vk}\Big[\frac{f(\Omega_{\vk}^+)-f(\Omega_{\vk}^-)}{\Delta \Omega_{\vk}}\Big]J_1\sin^2{\ah_1}\gamma_{1,\vk}^2+1\Bigg\}\,,
\end{align}
\begin{align}
\frac{\partial F}{\partial \Phi_{\vq}}&=2J_1\sin^2{\ah_1}|\Phi_{\vq}|\notag\\
&\times\Bigg\{16\sum_{\vk}\Big[\frac{f(\Omega_{\vk}^+)-f(\Omega_{\vk}^-)}{\Delta \Omega_{\vk}}\Big]J_1\sin^2{\ah_1}\gamma_{1,\vk+\vq/2}^2+1\Bigg\}\,,
\end{align}
\begin{align}
\frac{\partial F}{\partial S_{\vqaf}}&=2J_{\vqaf}S_{\vqaf}\notag\\
&\times\Bigg\{\sum_{\vk}\Big[\frac{f(\Omega_{\vk}^+)-f(\Omega_{\vk}^-)}{\Delta \Omega_{\vk}}\Big]J_{\vqaf}-1\Bigg\}\,,
\end{align}
\begin{align}
\frac{\partial F}{\partial \lambda_0}&=\sum_{\vk}\Big[f(\Omega_{\vk}^+)+f(\Omega_{\vk}^-)\Big]-1\,.
\end{align}
where $f(\omega)\equiv \frac{1}{1+\exp\beta\omega}$ denotes the Fermi function, and 
$\Delta\Omega_\vk\equiv \Omega_{\vk}^+-\Omega_{\vk}^-=2\sqrt{
(J_{\vqaf})^2\vert S_{\vq_{\text{AF}}}\vert^2+16(J_1^{\text{SL}})^2\big[(\gamma_{1,\vk})^2\vert\Phi_{1}\vert^{2}+(\gamma_{\vq,\vk})^2\vert\Phi_{\vq}\vert^{2}\big]
}$.
In the following it will be convenient to introduce the field-dependent sums: 
\begin{align}
A_{\lambda_0}&\equiv\frac{1}{N}\sum_{\vk}\Big[f(\Omega_{\vk}^+)+f(\Omega_{\vk}^-)\Big]\,,
\label{DefAlambda0}\\
A_{\Phi_2}&\equiv\frac{1}{N}\sum_{\vk}\Big[f(\Omega_{\vk}^+)+f(\Omega_{\vk}^-)\Big]\gamma_{2\vk}\,,
\label{DefAPhi2}\\
A_{\Phi_1}&\equiv\frac{1}{N}\sum_{\vk}\frac{f(\Omega_{\vk}^+)-f(\Omega_{\vk}^-)}{\Delta\Omega_{\vk}}\gamma_{1\vk}^2\,,
\label{DefAPhi1}\\
A_{\Phi_{\vq}}&\equiv\frac{1}{N}\sum_{\vk}\frac{f(\Omega_{\vk}^+)-f(\Omega_{\vk}^-)}{\Delta\Omega_{\vk}}\gamma_{1\vk\vq}^2\,,
\label{DefAPhiQ}\\
A_{S_{\vq_{\text{AF}}}}&\equiv\frac{1}{N}\sum_{\vk}\frac{f(\Omega_{\vk}^+)-f(\Omega_{\vk}^-)}{\Delta\Omega_{\vk}}\,.  
\label{DefASQ}
\end{align}
After some standard algebra, the seven saddle point equations for $F(\alpha_1,\alpha_2,\lambda_0,\Phi_{1},\Phi_{\vq},\Phi_2,S_{\vq_{\text{AF}}})$ are rewritten as : 
\begin{align}
J_1\sin{2\ah_1}\Big(8\gamma_{1,\vqaf}|S_{\vqaf}|^2+|\Phi_1|^2+|\Phi_{\vq}|^2\Big)&=0\,,\label{eq:sa1}\\
J_2\sin{2\ah_2}\Big(|\Phi_2|^2+\gamma_{2,\vqaf}|S_{\vqaf}|^2\Big)&=0\,,\label{eq:sa2}\\
J_2\sin{\ah_2}\Big(A_{\Phi_2}+2|\Phi_2|\Big)&=0\,,\label{eq:sphi}\\
J_1|\Phi_1|\sin{\ah_1}\Big(16J_1\sin^2{\ah_1}A_{\Phi_1}+1\Big)&=0\,,\label{eq:sph0}\\
J_1|\Phi_{\vq}|\sin{\ah_1}\Big(16J_1\sin^2{\ah_1}A_{\Phi_{\vq}}+1\Big)&=0\,,\label{eq:sphq}\\
J_{\vqaf}S_{\vqaf}\Big(A_{S_{\vqaf}}J_{\vqaf}-1\Big)&=0\,,\label{eq:ssqaf}\\
A_{\lambda_0}&=1\,.\label{eq:slam}
\end{align}
These equations may have some trivial solutions that correspond to giving $\alpha_1$ and/or $\alpha_2$  
the extreme values $0$ and $\pi/2$. 
This leads to four various cases that are defined in table~\ref{Table1}. Hereafter, the system of saddle-point 
relations~(\ref{eq:sa1}, \ref{eq:sa2}, \ref{eq:sphi}, \ref{eq:sph0}, \ref{eq:sphq}, \ref{eq:ssqaf}, \ref{eq:slam}) 
is rewritten accordingly to the simplifications provided by each case. In all cases, we still have to solve the saddle point equation for the Lagrange multiplier $\lambda_0$: 
\begin{align}
A_{\lambda_0}&=1\,. \label{eq:slam2}
\end{align}
For the other fields we are thus left with: 

\subsection{Trivial solutions: {\bf Case A}}
Here we consider the trivial cases where both $\ah_1$ and $\ah_2$ take extreme values 
$\pi/2$ or $0$. There are naturally four possibilities that are analyzed sub-case by sub-case hereafter. 
Most of the saddle point equations are trivially satisfied, and we analyze here the relevant relations that still 
remain. 

\subsubsection{Sub-case \texorpdfstring{$(\ah_1,\ah_2)=(0,0)$}{(a1,a2)=(0,0)}}
This situation corresponds to the classical magnetic mean-field approximation. 
In this case, only magnetic order is considered, with the two possible ordering wave-vectors  
$\vqaf^{\rm I}$ and $\vqaf^{\rm II}$. The saddle point equation for $S_{\vqaf}$ and a given ordering wave-vector 
is: 
\begin{align}
J_{\vqaf}S_{\vqaf}\Big(A_{S_{\vqaf}}J_{\vqaf}-1\Big)&=0\,.
\end{align}

\subsubsection{Sub-case \texorpdfstring{$(\ah_1,\ah_2)=(\pi/2,0)$}{(a1,a2)=(p/2,0)}}
Here, the interplane spin liquid fields compete or coexist 
with the magnetic order originating from the inplane Weiss field $J_{2}^{\text{Weiss}}$. 
The saddle point equations for $\Phi_1$, $\Phi_\vq$, and $S_{\vqaf}$ are: 
\begin{align}
J_1|\Phi_1|\Big(16J_1A_{\Phi_1}+1\Big)&=0\,,\\
J_1|\Phi_{\vq}|\Big(16J_1A_{\Phi_{\vq}}+1\Big)&=0\,,\\
J_2\gamma_{2,\vqaf}S_{\vqaf}\Big(2J_2\gamma_{2,\vqaf}A_{S_{\vqaf}}-1\Big)&=0\,.
\end{align}

\subsubsection{Sub-case \texorpdfstring{$(\ah_1,\ah_2)=(0,\pi/2)$}{(a1,a2)=(0,p/2)}}
Here, the different layers in $(a,b)$ directions are decoupled from each other in a pure Weiss field channel. 
Inside each layer, the mean-field decoupling is purely spin-liquid. 
The saddle point equations for $\Phi_2$ and $S_{\vqaf}$ are: 
\begin{align}
J_2\Big(A_{\Phi_2}+2|\Phi_2|\Big)&=0\,,\\
J_1\gamma_{1,\vqaf}S_{\vqaf}\Big(8J_1\gamma_{1,\vqaf}A_{S_{\vqaf}}-1\Big)&=0\,.
\end{align}

\subsubsection{Sub-case \texorpdfstring{$(\ah_1,\ah_2)=(\pi/2,\pi/2)$}{(a1,a2)=(p/2,p/2)}}
This corresponds to a pure spin liquid state with interplane fields $\Phi_1$, $\Phi_\vq$, and inplane 
field $\Phi_2$. The saddle point equations are: 
\begin{align}
J_2\Big(A_{\Phi_2}+2|\Phi_2|\Big)&=0\,,\\
J_1|\Phi_1|\Big(16J_1A_{\Phi_1}+1\Big)&=0\,,\\
J_1|\Phi_{\vq}|\Big(16J_1A_{\Phi_{\vq}}+1\Big)&=0\,. 
\end{align}


\subsection{Case B}
Here we consider that $\alpha_1$ is fixed to an extreme value ($0$ or $\pi/2$), and  
$\alpha_2$ is a free parameter. Since extremal values of $\alpha_2$ have been already considered in case A, we thus assume the strict inequality $0<\alpha_2<\pi/2$. 
Eq.~(\ref{eq:sa2}) can thus be simplified as: 
\begin{align}\label{eq:cb}
|\Phi_2|^2+\gamma_{2,\vqaf}|S_{\vqaf}|^2&=0\,,
\end{align}
Putting aside the trivial solution with vanishing fields, this relation requires an ordering wave-vector such that $\gamma_{2,\vqaf}<0$. Invoking the definition Eq.~(\ref{defgamma2k}), we check easily that $\gamma_{2,\vqaf^{\rm II}}=-2$ and $\gamma_{2,\vqaf^{\rm I}}=+2$. Therefore 
we consider only the ordering wave-vector $\vqaf^{II}=(1/2,1/2,0)$ for this case. 
Eq.~(\ref{eq:cb}) enforces linearity between the fields: 
\begin{align}
\vert\Phi_2\vert=\vert S_{\vqaf}\vert\sqrt{2}~.\label{eq:phi2eqSQ}
\end{align}
This relation and Eq.~(\ref{eq:slam2}) have to be completed by the other relevant saddle point equations that are rewritten  as follows: 
\begin{align}
J_2\Big(A_{\Phi_2}+2|\Phi_2|\Big)&=0\,,\\
J_2S_{\vqaf}\Big[4J_2\cos^2{(\ah_2)}A_{S_{\vqaf}}+1\Big]&=0\,, 
\end{align}
and also: 
\subsubsection{Sub-case \texorpdfstring{$\ah_1=0$}{a1=0}:}
\begin{align}
\Phi_1=\Phi_\vq&=0\,. 
\end{align}
\subsubsection{Sub-case \texorpdfstring{$\ah_1=\pi/2$}{a1=p/2}:}
\begin{align}
J_1|\Phi_1|\Big(16J_1A_{\Phi_1}+1\Big)&=0\,,\\
J_1|\Phi_{\vq}|\Big(16J_1A_{\Phi_{\vq}}+1\Big)&=0\,.
\end{align}

\subsection{Case C}
This case corresponds to $\sin{(2\ah_2)}=0$ and a strict inequality $0<\alpha_1<\pi/2$. 
Here we first consider Eq.~(\ref{eq:sa1}), that is rewritten as: 
\begin{align}\label{eq:sa1casc}
8\gamma_{1,\vqaf}|S_{\vqaf}|^2+|\Phi_1|^2+|\Phi_{\vq}|^2&=0\,.
\end{align}
Excluding the trivial solution with all fields vanishing, the AF ordering wave-vector must satisfy 
$\gamma_{1,\vqaf}<0$. Invoking the definition Eq.~(\ref{defgamma1k}), we check easily that $\gamma_{1,\vqaf^{\rm I}}=-1$ and $\gamma_{1,\vqaf^{\rm II}}=+1/2$. Therefore 
we consider only the ordering wave vector $\vqaf^{I}=(1,1,1)$ for this case, and  Eq.~(\ref{eq:sa1casc}) reads: 
\begin{align}\label{eq:sa1cascbis}
8|S_{\vqaf}|^2=|\Phi_1|^2+|\Phi_{\vq}|^2~.
\end{align}
In case C, this relation, together with Eq.~(\ref{eq:slam2}) has to be completed by the following  relevant saddle point equations: 
\begin{align}
J_1|\Phi_1|\Big(16J_1A_{\Phi_1}\sin^2{\ah_1}+1\Big)&=0\,,\\
J_1|\Phi_{\vq}|\Big(16J_1A_{\Phi_{\vq}}\sin^2{\ah_1}+1\Big)&=0\,.
\end{align}
and also: 
\subsubsection{Sub-case \texorpdfstring{$\ah_2=0$}{a2=0}:}
\begin{align}
S_{\vqaf}\Big[A_{S_{\vqaf}}\big(8J_1\cos^2{\ah_1}-4J_2\big)+1\Big]&=0\,, \\
\Phi_2&=0\,.
\end{align}
\subsubsection{Sub-case \texorpdfstring{$\ah_2=\pi/2$}{a2=p/2}:}
\begin{align}
J_2\Big(A_{\Phi_2}+2|\Phi_2|\Big)&=0\,,\\
S_{\vqaf}\Big(8J_1A_{S_{\vqaf}}\cos^2{\ah_1}+1\Big)&=0\,.
\end{align}

\subsection{Case D}
This case is in principle the most general one, where both $\alpha_1$ and $\alpha_2$ are considered as free parameters. Since extreme values $0$ or $\pi/2$ have already been considered in previous cases, we assume here strict equalities 
$0<\alpha_1<\pi/2$ and $0<\alpha_2<\pi/2$. Therefore, Eqs.~(\ref{eq:sa1}) 
and~(\ref{eq:sa2}) can be simplified as: 
\begin{align}\label{eq:casd}
8\gamma_{1,\vqaf}|S_{\vqaf}|^2+|\Phi_1|^2+|\Phi_{\vq}|^2&=0\,,\\
|\Phi_2|^2+\gamma_{2,\vqaf}|S_{\vqaf}|^2&=0\,. 
\end{align}
The only way to obtain a solution without all fields vanishing would require at least 
$S_{\vqaf}\neq 0$. The corresponding AF ordering wave-vector $\vqaf$ would have to satisfy 
both $\gamma_{1,\vqaf}<0$ and $\gamma_{2,\vqaf}<0$. 
Nevertheless, invoking definitions~(\ref{defgamma1k}) and~(\ref{defgamma2k}), we check easily that $\gamma_{1,\vqaf^{\rm I}}<0$ but $\gamma_{2,\vqaf^{\rm I}}>0$, and 
$\gamma_{1,\vqaf^{\rm II}}>0$ but $\gamma_{2,\vqaf^{\rm II}}<0$. 
We thus conclude that neither $\vqaf^{\rm I}$ nor $\vqaf^{\rm II}$ ordering wave-vectors can lead to such a solution.

\bibliography{biblio_j1j2j3_BCT}

\end{document}